\documentclass[10pt,twocolumn,superscriptaddress,floats,showpacs,nobalancelastpage,longbibliography,prb]{revtex4-2}

\usepackage{array,longtable}
\newcolumntype{L}{>{\tiny $}p{0.33\columnwidth}<{$}}
\newcolumntype{M}{>{\scriptsize $}p{0.33\columnwidth}<{$}}
\newcolumntype{N}{>{\scriptsize $}p{0.43\columnwidth}<{$}}
\usepackage{amsmath}
\usepackage{amssymb}
\usepackage{amsfonts}
\usepackage{graphicx}
\usepackage{hyperref}
\usepackage{tabularx}
\usepackage{wasysym}

\usepackage{float}
\usepackage{graphics}
\usepackage[caption=false]{subfig}
\usepackage{tikz}
\usepackage{times}
\usepackage{dsfont}
\usepackage{color}
\usepackage{subfig}

\usetikzlibrary{calc}
\tikzset{
    ncbar angle/.initial=90,
    ncbar/.style={
        to path=(\tikztostart)
        -- ($(\tikztostart)!#1!\pgfkeysvalueof{/tikz/ncbar angle}:(\tikztotarget)$)
        -- ($(\tikztotarget)!($(\tikztostart)!#1!\pgfkeysvalueof{/tikz/ncbar angle}:(\tikztotarget)$)!\pgfkeysvalueof{/tikz/ncbar angle}:(\tikztostart)$)
        -- (\tikztotarget)
    },
    ncbar/.default=0.5cm,
}
\tikzset{square left brace/.style={ncbar=0.5cm}}
\tikzset{square right brace/.style={ncbar=-0.5cm}}

\usepackage[normalem]{ulem}

\allowdisplaybreaks[1]

\newif\ifhyper
\hypertrue
\ifhyper
\hypersetup{
   citecolor = {green},
   colorlinks = {true}, 
   urlcolor = {blue}, 
   linkcolor = {blue}
}
\fi

\newenvironment{diagram}
{
\begin{tikzpicture}[baseline = (X.base),every node/.style={scale=0.8},scale=0.45]
}
{
\end{tikzpicture}
}

\begin{document}
\title{Cubic ferromagnet and emergent $U(1)$ symmetry on its phase boundary}

\author{Wei-Lin Tu}
\email{weilintu@keio.jp}
\affiliation{Division of Display and Semiconductor Physics, Korea University, Sejong 30019, Korea}
\affiliation{Faculty of Science and Technology, Keio University, 3-14-1 Hiyoshi, Kohoku-ku, Yokohama 223-8522, Japan}

\author{Xinliang Lyu}
\affiliation{Institute for Solid State Physics, University of Tokyo, Kashiwa, Chiba 277-8581, Japan}

\author{S. R. Ghazanfari}
\affiliation{Institute for Solid State Physics, University of Tokyo, Kashiwa, Chiba 277-8581, Japan}

\author{Huan-Kuang Wu}
\affiliation{Department of Physics, Condensed Matter Theory Center and Joint Quantum Institute, University of Maryland, College Park, MD 20742, USA}

\author{Hyun-Yong Lee}
\email{hyunyong@korea.ac.kr}
\affiliation{Division of Display and Semiconductor Physics, Korea University, Sejong 30019, Korea}
\affiliation{Department of Applied Physics, Graduate School, Korea University, Sejong 30019, Korea}
\affiliation{Interdisciplinary Program in E$\cdot$ICT-Culture-Sports Convergence, Korea University, Sejong 30019, Korea}

\author{Naoki Kawashima}
\email{kawashima@issp.u-tokyo.ac.jp}
\affiliation{Institute for Solid State Physics, University of Tokyo, Kashiwa, Chiba 277-8581, Japan}
\affiliation{Trans-scale Quantum Science Institute, The University of Tokyo, Bunkyo, Tokyo 113-0033, Japan}

\date{\today}

\begin{abstract}

We study the simplest quantum lattice spin model for the two-dimensional~(2D) cubic ferromagnet by means of mean-field analysis and tensor network calculation.
While both methods give rise to similar results in detecting related phases, the 2D infinite projected entangled-pair state~(iPEPS) calculation provides more accurate values of transition points.
Near the phase boundary, moreover, our iPEPS results indicate that it is more difficult to pin down the orientation of magnetic easy axes, and we interpret it as the easy-axis softening.
This phenomenon implies an emergence of continuous $U(1)$ symmetry, which is indicated by the low-energy effective model and has been analytically shown by the field theory.
Our model and study provide a concrete example for utilizing iPEPS near the critical region, showing that the emergent phenomenon living on the critical points can already be captured by iPEPS with a rather small bond dimension.
\end{abstract}

\maketitle

\section{\label{Intro}Introduction}
For magnetic materials, as the temperature drops below the Curie point, spins as the microscopic objects amount to a collective behavior and ferromagnetism forms. 
In the most general picture, where spins enjoy the unitary transformation in three dimensions, formation of magnetic order stands for the breaking of $O(3)$ symmetry group. 
However, due to many possible reasons such as the lattice structure, within real-world materials the ferromagnetic moment tends to align along certain directions, the so-called easy axes. 
When the easy-axis orientation follows the principal or diagonal direction of the cubic~(isometric) crystal structure, they are referred to as the cubic ferromagnets~\cite{PhysRev.52.1178, IntroMag}.
In fact, no matter being weak or strong it is a very common feature for the magnetic materials and thus affects the microscopic mechanism.

Recently, layered magnetic materials have drawn researchers' attention because of their broad range of potential applications within the two-dimensional~(2D) layers~\cite{nrevphys1}. 
Among them, the ferromagnetic semiconductor thin film is of special interest because of its promising features from industrial point of view~\cite{nat.mat.9.965}, and various magnetic anisotropies can also be found in these materials~\cite{RevModPhys.86.187}. 
In many related works upon such magnetic materials, the determination of easy-axis alignment plays an important role in their studies to explain the experimental observation. 
However, the introduction of quantum effect and its influence in the low temperature are seldom discussed because most studies followed the mean-field paradigm. 
Moreover, despite a great amount of effort for its theoretical understanding through Zener's or other phenomenological models~\cite{RevModPhys.86.187}, the corresponding microscopic picture is often overlooked.

The first effective Landau theory for cubic systems appeared in Ref.~\cite{PhysRevB.1.3599}, and an effective lattice model with single-ion anisotropy can be constructed, leading to preferable magnetic easy axes after symmetry breaking~\cite{PhysRevB.23.4556}.
It has later been shown by Sznajd and Domański that the Landau free energy of this many-body Hamiltonian leads to the corresponding Landau-Ginzburg-Wilson~(LGW) Hamiltonian being the $n$-vector model with a cubic anisotropy~\cite{J.Magn.Magn.Mater.42, Phys.StatusSolidiB129}.
This simplest lattice model with cubic anisotropy is free from the vicious sign problem, but remains hard to probe with quantum Monte Carlo~(QMC) due to an unexpected freezing of local spin moments, which we elucidate in Appendix~\ref{qmc}. 
To the best of our knowledge, there has been no elaborated result beyond the mean-field studies~\cite{J.Magn.Magn.Mater.42, Phys.StatusSolidiB129, Phys.StatusSolidiB135, DOMANSKI1988306, PhysRevB.59.4176} so far for this lattice model, despite some more recent studies showing the possible existence of quadrupolar phase with easy axes aligning along $\langle 100 \rangle$ in one dimension by utilizing the perturbation theory and density-matrix renormalization group~(DMRG)~\cite{Eur.Phys.J.B.5, PhysRevB.59.13764, Eur.Phys.J.B.17}.

In this work, we study the quantum many-body model with both the mean-field analysis and the 2D tensor network ansatz in the thermodynamic limit, the infinite projected entangled-pair states~(iPEPS)~\cite{PhysRevLett.101.250602, NatRevPhys.1.538, RevModPhys.93.045003}, in order to take into account the quantum entanglement among sites, which goes beyond the mean-field treatment, for uncovering its more precise phase diagram.
Moreover, near the phase boundary the canted magnetic order is susceptible in its orientation according to iPEPS results, suggesting the disappearance of easy axes and interpreted as the easy-axis softening.
Since the perturbative renormalization group~(RG) and other methods have revealed that the continuous symmetry emerges for a two-component phase transition~\cite{PhysRevB.8.4270, PhysRevB.61.15136, PhysRevB.84.125136, ADZHEMYAN2019332, PhysRevD.104.105013}, the easy-axis softening by iPEPS indicates such emergence of $U(1)$ symmetry on a lattice model.
Our results also consolidate the usage of iPEPS near the critical area because even with a rather small bond dimension the predicted emergent phenomenon can still be clearly detected, which is, to our best knowledge, demonstrated for the very first time.

\section{Lattice Model with Cubic Anisotropy}

\subsection{\label{model}Hamiltonian}
As mentioned in the Introduction, in this work we are interested in the cubic ferromagnetic thin films and its simplest many-body Hamiltonian in the square lattice is~\cite{DOMANSKI1988306}
\begin{equation}
\begin{aligned}
H&=-J\sum_{\langle i,j \rangle}\vec{S}_i \cdot \vec{S}_j+K\sum_{i,\alpha}(S^\alpha_i)^4 -h\sum_{i}S^z_i\\
&=H_J+H_{K-h},
\end{aligned}
\label{Hamil}
\end{equation}
with $S=2$ and $\alpha=x,y,z$. We divide the Hamiltonian into $H_J$ and $H_{K-h}$ for later convenience.
Spins on nearest-neighbor sites~($\langle i,j\rangle$) are connected through a ferromagnetic Heisenberg interaction~($J>0$). The second term reflects the cubic anisotropy and with $K>0$, its easy axes go along $\langle 111 \rangle$.
In this work we assign $K=1$ as the energy unit.
For a 2D magnet, especially the semiconductor thin film, frequently it functions under an external magnetic field. Thus, a Zeeman term along $z$-axis is also considered when a magnetic field is present. 
Note that Eq.~(\ref{Hamil}) does not commute with the total-spin operator: $[H,\sum_iS^z_i]\neq 0$, and thus it does not possess the spin~(or particle) conservation symmetry, meaning that the additional phase factor introduced by $U(1)$ transformation will alter the Hamiltonian.
Furthermore, with a non-zero $h$ that deforms the original cubic lattice into a tetragonal one, we have only four favored directions for the magnetization lying above the $x-y$ plane.
In fact, the symmetry group of our model can be probed by replacing $(\pm S^\beta,\pm S^{\bar{\beta}})$ with $\beta,\bar{\beta} =x,y$ and $\beta\neq\bar{\beta}$ into Eq.~(\ref{Hamil}) while leaving itself unchanged.
Such transformation includes 4 rotations and 4 reflections in the plane, and thus they compose a dihedral group with order 4 or written as $D_8$ in the abstract algebra.
By choosing $K=1$ in this work, four favored directions possess azimuthal angles $\theta$, shown in Fig.~\ref{fig1}(a), equal to $\pi/4,3\pi/4,5\pi/4$, and $7\pi/4$.
In what follows, we restrict $\theta$ to be in the range $\theta\in[0,\frac{\pi}{4}]$ without loss of generality.

\subsection{\label{onesite}The one-site ($J=0$) limit}

\begin{figure}[pbt]
\centering
	\includegraphics[width=1.0\columnwidth]{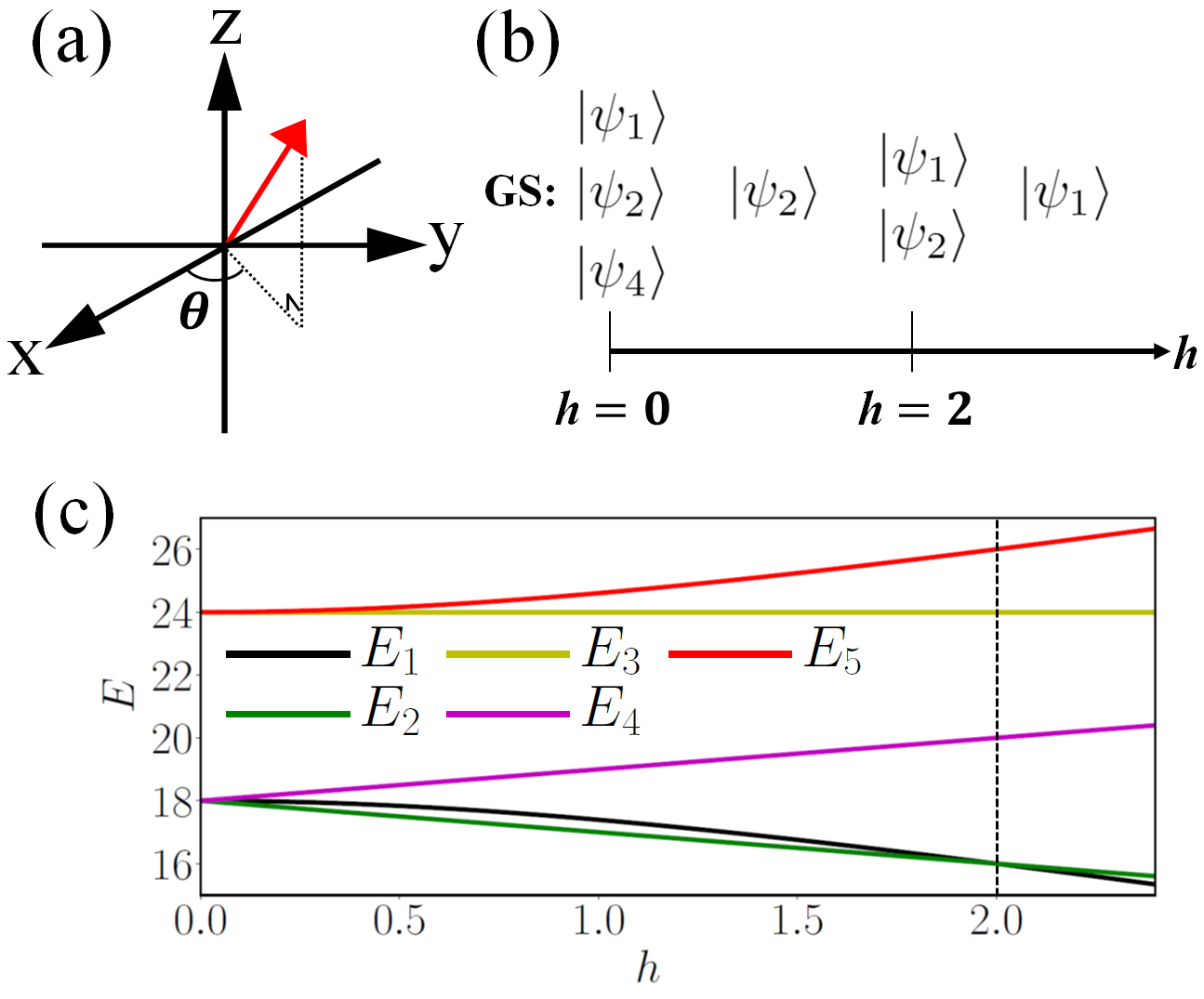}
\caption{(a) The spontaneous magnetization (the red arrow) and the angle $\theta$. When a canted phase appears, planar magnetic order points in a certain direction with relative angle $\theta$ to the $x$-axis ($[100]$). (b) Ground-state phase diagram for the single-site Hamiltonian ($J=0$). Definition for each state is recorded in Eq.~(\ref{basis}). (c) The eigen-energy for each local state along with $h$. Energies are calculated according to Eq.~(\ref{prefactor}) and (\ref{eigenvalues}). Dashed line at $h=2$ indicates the degenerate~(transition) point.}
\label{fig1}
\end{figure}

We begin our discussion from studying one special case as $J$ is chosen to be zero.
Under this choice, $H=H_{K-h}$ and it is a one-site Hamiltonian which can be directly solved. 
Because $S=2$, the local Hilbert space is five-dimensional and $S_z=1$, $0$, and $-1$ states are still good eigenstates. On the other hand, due to the quartic term $S_z=2$ and $-2$ states are connected and new eigenbasis needs to be formed. 
The results are
\begin{equation}
\begin{aligned}
\begin{bmatrix}
|\psi_1 \rangle \\
|\psi_2 \rangle \\
|\psi_3 \rangle \\
|\psi_4 \rangle \\
|\psi_5 \rangle 
\end{bmatrix}
=
\begin{bmatrix}
N(|2\rangle-\alpha |-2\rangle)\\
|1\rangle\\
|0\rangle\\
|-1\rangle\\
N(\alpha |2\rangle+|-2\rangle)
\end{bmatrix},
\end{aligned}
\label{basis}
\end{equation}
where $|2\rangle$, $|1\rangle$, $|0\rangle$, $|-1\rangle$, and $|-2\rangle$ represent the eigenstates of $S^z$ with
\begin{equation}
\begin{aligned}
&\alpha=\frac{\sqrt{4h^2+9}-2h}{3}\ (\leqslant 1),\;N=\frac{1}{\sqrt{\alpha^2+1}}.
\end{aligned}
\label{prefactor}
\end{equation}
The corresponding eigenvalues are
\begin{equation}
\begin{aligned}
\begin{bmatrix}
E_{1} \\
E_{2} \\
E_{3} \\
E_{4} \\
E_{5} 
\end{bmatrix}
=
\begin{bmatrix}
-3\alpha^{-1}+21+2h\\
18-h\\
24\\
18+h\\
3\alpha+21+2h
\end{bmatrix},
\end{aligned}
\label{eigenvalues}
\end{equation}
with $E_n$ being the eigen-energy of $|\psi_n\rangle$. We note that the formation of this ``reshuffled'' Hilbert basis results from the fact that the $K$ and $h$ terms do not commute.
Therefore, a new set of orthonormal basis is generated.
By varying $h$, we plot the ground state~(GS) phase diagram in Fig.~\ref{fig1}(b). 
There are two special points where degeneracy takes place. One is at $h=0$ with a three-fold degeneracy among $|\psi_1 \rangle$, $|\psi_2 \rangle$, and $|\psi_4 \rangle$. 
The other lies at $h=2$ with a two-fold degeneracy between $|\psi_1 \rangle$ and $|\psi_2 \rangle$. 
We will especially focus on the vicinity of the second degenerate point, $h=2$.
There, a transition of the two-level system takes place, meaning that a first-order transition happens with a sudden jump of the order parameter $\langle S^z \rangle$, with $\langle \psi_1|S^z|\psi_1 \rangle=1.6$ and $\langle \psi_2|S^z|\psi_2 \rangle=1$ where $S^z=\frac{1}{N_L}\sum_iS^z_i$ and $N_L$ stands for the number of lattice sites.
%

We also plot the eigen-energies~(Eq.~(\ref{eigenvalues})) along $h$ for each state in Fig.~\ref{fig1}(c). 
One obvious feature lies in the fact that as $h$ is strong enough, $|\psi_3 \rangle$, $|\psi_4 \rangle$, and $|\psi_5 \rangle$ are largely gapped from the rest two states.
Moreover, $|\psi_1 \rangle$ and $|\psi_2\rangle$ share very close energies around $h=2$. The above observations suggest that near the transition point, other higher-energy states can only serve as the ``tiebreaker'' through quantum fluctuation after we turn on $J$.
This will become an important feature in the following discussion.

\subsection{\label{Jcoup}Inclusion of Heisenberg coupling}

We will now study the effect while including the $J$ term. 
As mentioned in the previous section, the Heisenberg term serves as the tunneling from low-energy states to higher ones.
It will also introduce the inter-site correlation that leads to a collective behavior of local spins, forming the ferromagnetic state.
Our remaining task is to investigate all possible phases after $J$ is turned on and try to probe the critical behavior on the phase boundary.

Initially, we can first try to picture what kinds of phases could emerge. From the information of $H_{K-h}$, we have learned that there is a two-level system and both states are gapped.
Therefore, they will remain until $J$ is strong enough to overcome the energy difference between $|\psi_1 \rangle$ and $|\psi_2 \rangle$, thus triggering a phase transition. 
Since both are polarized in the $z$-direction, we call them the polarized states, $P_1$ and $P_2$, coming from $|\psi_1\rangle$ and $|\psi_2\rangle$ respectively. 
Note that although from the symmetry point of view both $P$ phases are identical, because of the two-level nature we label them with different indices, same as the fact that we call the phase of matter with translational invariance possessing a larger~(smaller) density the liquid~(gas) phase.
Besides the polarized phase as $J$ is strong enough the condensation takes place, leading to the off-diagonal magnetic moments. 
Therefore, we expect that a canted~($C$) phase, whose magnetization no longer aligns along the $z$-axis, should also appear in the phase diagram.
The transition between $P$ and $C$ phases then becomes the central issue in the following content.

\section{Methods and Results}

\subsection{\label{MF}Mean-field analysis}

\begin{figure}[pbt]
\centering
	\includegraphics[width=1.0\columnwidth]{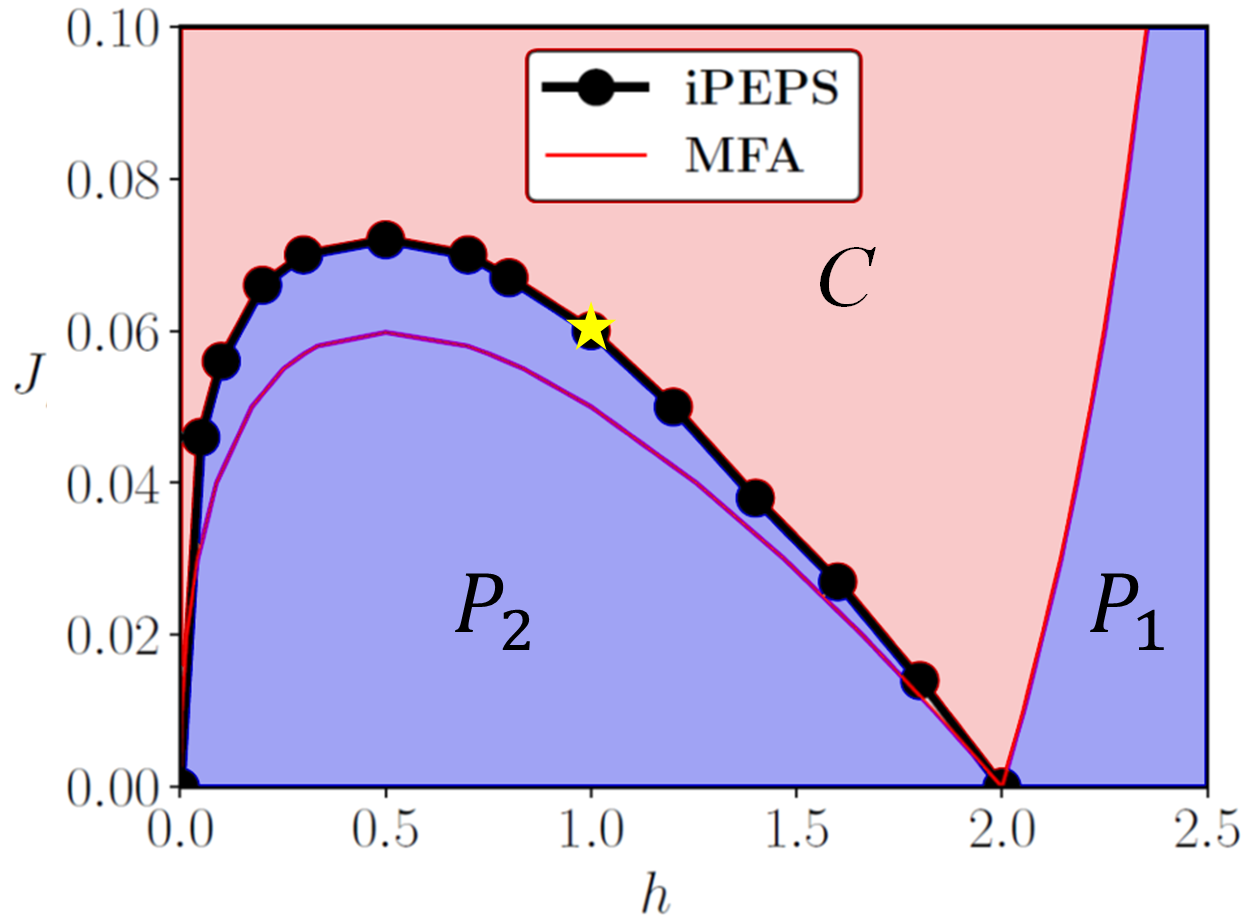}
\caption{The phase diagram by MFA~(red thin lines) and iPEPS~(circle symbols) with $D=4$~($\chi=70$). For MFA, phase boundaries are obtained by self-consistently solving Eq.~(\ref{HamilMF}). Two polarized~($P_1$ and $P_2$) phases are connected through $h=2$ while canted~($C$) phase forms when $J$ comes into play. The yellow star indicates the transition point probed by iPEPS with $(D,\chi)=(5,50)$. See Section~\ref{ipeps} and Appendix~\ref{scaling} for detailed discussion.}
\label{fig2}
\end{figure}

We first study the full Hamiltonian, Eq.~(\ref{Hamil}), with mean-field approximation~(MFA), sometimes also referred to as the molecular field approximation. The mean-field Hamiltonian reads
\begin{equation}
\begin{aligned}
H_{\text{MF}}=&-J\sum_{\langle i,j \rangle}\vec{S}_i \cdot \langle \vec{S}_j\rangle+\sum_{i,\alpha}(S^\alpha_i)^4 -h\sum_{i}S^z_i.
\end{aligned}
\label{HamilMF}
\end{equation}
By introducing the mean-field of $\langle \vec{S}_j\rangle$ Eq.~(\ref{HamilMF}) again reduces to a one-site Hamiltonian and can be solved iteratively by assuming the local state to be
\begin{equation}
\begin{aligned}
|\psi_{\rm local}\rangle=c_0|\psi_1\rangle+\sum_{m=1}^{4}c_me^{i\phi_m}|\psi_{m+1}\rangle,
\end{aligned}
\label{localstate}
\end{equation}
where $c_0=\sqrt{1-\sum_mc^2_m}$.
Then, we variationally optimize the parameter set $\{c_m,\phi_m\}$ in lowering the energy, $\langle H_\text{MF}\rangle$.
To diagnose $P$ and $C$ phases, we rely on the order parameter
\begin{equation}
\begin{aligned}
M_p=\sqrt{\langle S^x\rangle^2+\langle S^y\rangle^2},
\end{aligned}
\label{order}
\end{equation}
where $M_p$ is nonzero in $C$ phase.
We emphasize that in $C$ phase, there is a four-fold degeneracy for its planar magnetization $M_p$, pointing along the diagonal direction in the x-y plane.
Without loss of generality, we adopt the magnetization with $\theta=\pi/4$ in the following discussion.
Such degeneracy has also been indicated in previous mean-field studies~\cite{DOMANSKI1988306}.
We plot the mean-field phase diagram in Fig.~\ref{fig2} as thin red lines, along with the phase boundary by iPEPS which will be discussed in the later sections.
Besides $M_p$, another relevant observable is $M_z=\langle S^z\rangle$. Recall that at $J=0$, we have a sudden jump of $M_z$ at $h=2$, indicating the first-order transition.
Because of the ferromagnetic nature, we can adopt the translational invariance in our calculation and thus $\langle S^\alpha \rangle=\langle S^\alpha_i \rangle$.
For the $P_2$ phase $M_z\simeq 1$ while $M_z> 1.6$ for the $P_1$ phase beyond $h=2$~(cf. $\langle \psi_1|S^z|\psi_1 \rangle=1.6$ and $\langle \psi_2|S^z|\psi_2 \rangle=1$ at $h=2$). 
Notice that $M_z$ does not have to be strictly equal to 1 because it is not a good quantum number. 

We emphasize two points here: (1) $M_z$ is not an order parameter because in all phases it is non-zero.
And (2) when $h=0$ the cubic symmetry is restored with 8 preferable directions of magnetization. In fact, it can be sensed from Eq.~(\ref{basis}) and (\ref{eigenvalues}) that when $h=0$, we have a three-fold degeneracy among $|1\rangle$, $|-1\rangle$, and $|\psi_1\rangle$ which is equal to $\frac{1}{\sqrt{2}}(|2\rangle-|-2\rangle)$.
In our study we focus on the scenario when $h\neq 0$; therefore, inside the $C$ phase $M_p$ always favors one of the tetragonal directions, which is consistent with previous results~\cite{DOMANSKI1988306}.

If we take a look at the optimized parameter set, $\{c_m,\phi_m\}$, we can see that the coefficient of a higher-energy $|\psi_m\rangle$ is proportional to $J/(E_m-E_g)$, where $E_m$ is its corresponding eigen-energy and $E_g$ is the ground state energy at $J=0$: $E_g=E_1$~($E_2$) at $h>2$~($<2$). 
Therefore, we can learn from Fig.~\ref{fig1}(c) that as $J$ remains relatively small, the coefficients of $|\psi_1\rangle$ and $|\psi_2\rangle$ should be dominant.
Moreover, the phase factors, $\phi_1$ to $\phi_4$, provide extra degrees of freedom in lowering $\langle H_\text{MF}\rangle$.
According to our MFA, in $C$ phase $(\phi_1,\phi_2,\phi_3,\phi_4)=(5\pi/4,3\pi/2,7\pi/4,0)$, suggesting that a coherence is attained and it leads to the $\theta=\pi/4$ magnetization.
It is crucial to note that such coherence is a consequence of our reshuffled local Hilbert space because in the sole Heisenberg ferromagnets there is no easy axis after the spontaneous symmetry breaking.

By far the physical scenario we have found agrees quite well with our intuitive assumption. The next question is if we can exploit a better numerical tool in order to provide a more accurate phase diagram.

\subsection{\label{ipeps}iPEPS study}

\begin{figure}[pbt]
\centering
	\includegraphics[width=1.0\columnwidth]{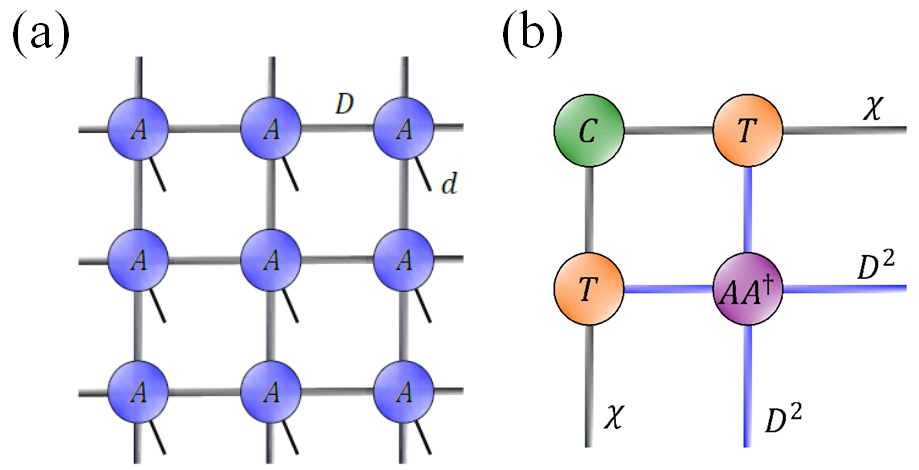}
\caption{(a) iPEPS scheme with rank-5 tensors on each lattice site where thick and thin solid lines represent virtual~($D$) and physical~($d$) bonds respectively. (b) Corner~($C$), edge~($T$) and double layer~($AA^\dagger$) tensors in the upper-left corner. Grey bonds possess dimension equal to $\chi$ which in general should be no smaller than $D^2$, and the dimension for blue bonds is $D^2$.}
\label{fig3}
\end{figure}

With a frustration-free many-body spin Hamiltonian, QMC is usually the prior option for conducting numerical studies.
However, due to the difficulty that we elucidate in Appendix~\ref{qmc}, it is unlikely to numerically solve Eq.~(\ref{Hamil}) by applying QMC. 
Therefore, we need to seek for another numerical tool and iPEPS will be applied for our purpose.
The iPEPS is a variational tensor network ansatz for approximating the ground state of a two dimensional quantum systems in the thermodynamic limit~\cite{PhysRevLett.101.250602, NatRevPhys.1.538, RevModPhys.93.045003}, in order to study our system beyond MFA.
Obeying the area law~\cite{ORUS2014117}, absence of sign problem and lattice size augmentation put iPEPS among one of the most desirable computational methods in many-body physics especially for strongly correlated systems. 
The basic idea in iPEPS consists of considering a repeating unit cell of interconnected tensors, so called the bulk tensors, and simulating an effective environment by constructing a series of border tensors. 
Each bulk tensor encodes the entanglement with its neighboring sites by having four virtual bonds with bond dimension equal to $D$, along with one physical index which reflects the local Hilbert space dimension~($d=5$ in this work).
Since we consider a ferromagnetic system in a square lattice, it then has a $1\times1$ repeating unit cell as shown in Fig.~\ref{fig3}(a).

To compute the norm and the observables we then contract the tensor~($A$) to its complex conjugate~($A^\dagger$) and trace out the physical bond so that it results in a double-layer tensor~($AA^\dagger$). 
This tensor object becomes the basic block while we approximate the environment tensors through the corner transfer matrix renormalization group~(CTMRG) procedure~\cite{doi:10.1143/JPSJ.65.891, PhysRevB.80.094403, PhysRevLett.113.046402}. 
Since the computational cost for the exact contraction of a 2D network exponentially grows along with the size, we construct the projector tensors through truncation after singular-value decomposing a tensor bond~\cite{PhysRevLett.113.046402}. 
Once the CTMRG converges, we obtain the corner $C$ and edge $T$ tensors as depicted in Fig.~\ref{fig3}(b). 
By constructing the environment tensors around the bulk tensor, we can extrapolate the system size to the thermodynamic limit and calculate the energy or desired physical observables.

\begin{figure*}[pbt]
\centering
	\includegraphics[width=1.8\columnwidth]{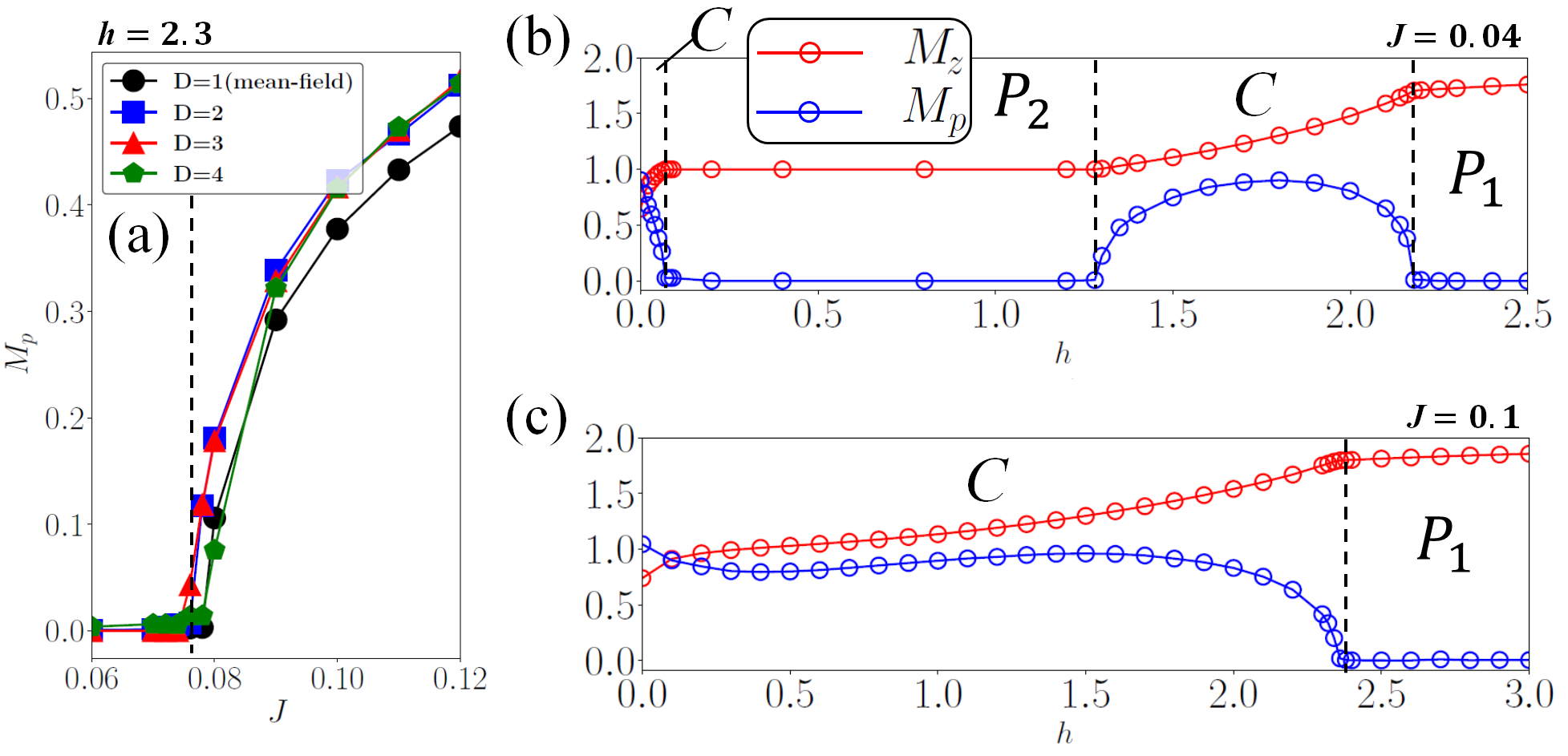}
\caption{iPEPS calculations~($D=4$ and $\chi=20$) for (a) The order parameter~($M_p$) probed by different bond dimensions along a vertical cut at $h=2.3$, and the plots of $M_z$ and $M_p$ observables along with $h$ for (b) $J=0.04$ and (c) $J=0.1$. A $C\to P_2 \to C\to P_1$ phase transition can be clearly seen in (b), while only two phases are detected in (c).}
\label{fig4}
\end{figure*} 

By construction iPEPS is suitable for studying gapped phases in two dimensions, because they fulfill the area law, and the fast decay of singular values makes the truncation during CTMRG reasonable.
On the other hand, whether iPEPS can still serve as an accurate ansatz for gapless phases is under debate.
Although we already have some solid examples, such as the expression of gapless Kitaev spin liquid using iPEPS~\cite{PhysRevLett.123.087203}, whether we can generally construct such structure for critical phases is uncertain.
Despite some recent works demonstrating how to conduct proper scaling with respect to the finite correlation length~\cite{PhysRevX.8.031030, PhysRevX.8.031031, PhysRevX.8.041033, PhysRevLett.129.200601}, a simple but direct diagnosis of critical behavior is not easy to obtain.
We will show, however, that our model along with the iPEPS results provide an easy yet desirable example showing the utility of this 2D tensor network ansatz even for the critical phases.

As a variational ansatz, various optimization methods based on the imaginary time evolution~(ITE) such as simple~\cite{PhysRevLett.101.250602}, full~\cite{PhysRevLett.101.090603} and fast-full updates~\cite{PhysRevB.92.035142} have already been proposed and implemented for oftentimes. 
On the other hand, although it is not an easy task to evaluate the energy gradient of each variational parameter~\cite{PhysRevB.94.035133, PhysRevB.94.155123}, a breakthrough in optimizing iPEPS emerged by adopting the idea of automatic differentiation (AD)~\cite{PhysRevX.9.031041}, which has been employed for optimizing problems with a large number of parameters and already proved its efficiency in many neural network studies. 
An advanced optimization scheme combining both ITE and AD has also been proposed recently~\cite{PhysRevResearch.4.043153}.

In this work, we adopt the variational optimization using AD for our ansatz.
The objective in AD optimization is to record down the computation graph from initial bulk tensors to the final energy estimation~(see the Method Section in Ref.~\cite{Commun.Phys.5.130}). 
After completing one computation flow~(one epoch), the energy gradients are evaluated through the backward propagation; hence we can make use of these gradients to update the tensor elements~(variational parameters) with a desired degree~(learning rate), until a desirable convergence is achieved. 
We can then utilize the converged ansatz for further calculation of physical observables. 
In this work, we adopt a practical package, peps-torch~\cite{Hasik}, for our calculation. 
Previously, this package has demonstrated a very good capability for various spin systems such as frustrated Heisenberg antiferromagnet~\cite{10.21468/SciPostPhys.10.1.012, PhysRevX.12.031039}, chiral spin liquid~\cite{PhysRevLett.129.177201}, and novel quantum magnetism~\cite{Commun.Phys.5.130}, as well as the bosonic system~\cite{PhysRevA.102.053306}.

We then present the numerical results obtained by iPEPS, starting from the phase diagram.
The black circles in Fig.~\ref{fig2} pin down the estimated transition points by iPEPS with bond dimensions $(D,\chi)=(4,70)$.
To strengthen the reliability, we also provide one point probed with $(D,\chi)=(5,50)$, which is the upper limit of our machine's capacity, in yellow star symbol.
The estimated transition point for $D=5$ is equal to the one of $D=4$ within a small $J$ variance~($\Delta J=0.002$).
Thus, we believe that the phase boundary present here is already very close to that of $D\to \infty$.
The way of deciding transition points is explained in Appendix~\ref{scaling}.

As we can clearly see, the phase boundary of $P_2$ dome is quantitatively different from the one by MFA, suggesting an improvement after adopting iPEPS.
For the other boundary at $h>2$, on the other hand, MFA is already very accurate.
In Fig.~\ref{fig4}(a) we present the order parameter, $M_p$, obtained with different bond dimensions using iPEPS along the $h=2.3$ vertical cut.
All the different trials predict a transition point at around $J=0.076$~(dashed line), meaning that the iPEPS calculation with $D>1$ does not provide a better prediction.
It is not surprising because the second boundary serves as the saturation line, beyond which the product state of $|\psi_1\rangle$ is a well estimated ansatz for its GS even with a nonzero $J$.

In Fig.~\ref{fig4}(b) and (c), we demonstrate two horizontal cuts of relevant observables, $M_z$ and $M_p$, for $J=0.04$ and $J=0.1$.
As shown in Fig.~\ref{fig4}(b), $M_p$ is nonzero in the beginning for small $h$ and then becomes zero entering the $P_2$ phase dome. 
As the magnetic field is further enhanced, $M_p$ appears again in $C$ phase. Finally, the state gets saturated and $M_p$ disappears. 
Such phase transition corresponds to the re-entrant $C\to P_2 \to C\to P_1$ transition revealed in Ref.~\cite{DOMANSKI1988306}. 
When $J$ is large enough outside the $P_2$ phase dome, however, we end up with only one $C$ and $P_1$ phases, as demonstrated in Fig.~\ref{fig4}(c) for $J=0.1$.

\begin{table}
\centering
\begin{tabular}{ccccc}\hline\hline
            & \begin{tabular}[c]{@{}c@{}}~$E_\text{GS}$~\\\end{tabular} & ~~$\theta_c/\pi$~~& ~~$M_z$~~& ~~$M_p$~~   \\ 
\hline
~$(J,h)=(0.08,0.5)$~ & 17.31491  & 0.18845 & 1.00833 & 0.64239   \\
~\;\;\;\;\;\;\;\;\;\;\;\;\;\;\;~ & 17.31426 & 0.25130 & 1.00884 & 0.62537 \\
\hline
~$(J,h)=(0.08,0.7)$~ & 17.11102 & 0.15063 & 1.03210 & 0.70821  \\
~\;\;\;\;\;\;\;\;\;\;\;\;\;\;\;~ & 17.11022 & 0.24194 & 1.03301 & 0.68585  \\
\hline
~$(J,h)=(0.06,1.2)$~ & 16.65518 & 0.00018 & 1.08426 & 0.74180  \\
~\;\;\;\;\;\;\;\;\;\;\;\;\;\;\;~ & 16.65518 & 0.04029 & 1.08427 & 0.74277  \\
~\;\;\;\;\;\;\;\;\;\;\;\;\;\;\;~ & 16.65529 & 0.09014 & 1.08398 & 0.74496  \\
~\;\;\;\;\;\;\;\;\;\;\;\;\;\;\;~ & 16.65527 & 0.12205 & 1.08426 & 0.74618  \\
~\;\;\;\;\;\;\;\;\;\;\;\;\;\;\;~ & 16.65528 & 0.13893 & 1.08433 & 0.74967  \\
~\;\;\;\;\;\;\;\;\;\;\;\;\;\;\;~ & 16.65509 & 0.19642 & 1.08440 & 0.73746  \\
~\;\;\;\;\;\;\;\;\;\;\;\;\;\;\;~ & 16.65513 & 0.22317 & 1.08450 & 0.74337  \\
~\;\;\;\;\;\;\;\;\;\;\;\;\;\;\;~ & 16.65509 & 0.25318 & 1.08475 & 0.73936  \\
\hline\hline
\end{tabular}
\caption{We provide the detailed numbers of simulations for three points near the $P_2$ dome. Numbers for trials with and without the local-minimum issue are shown in the first and last rows for each point with $D=4$ and $\chi=20$. The energy difference among distinct ansatz for $(J,h)=(0.06,1.2)$ is especially less apparent, whose reason will be explained in Section~\ref{HCB}.}
\label{tab1}
\end{table}

\begin{figure*}[pbt]
\centering
	\includegraphics[width=2.0\columnwidth]{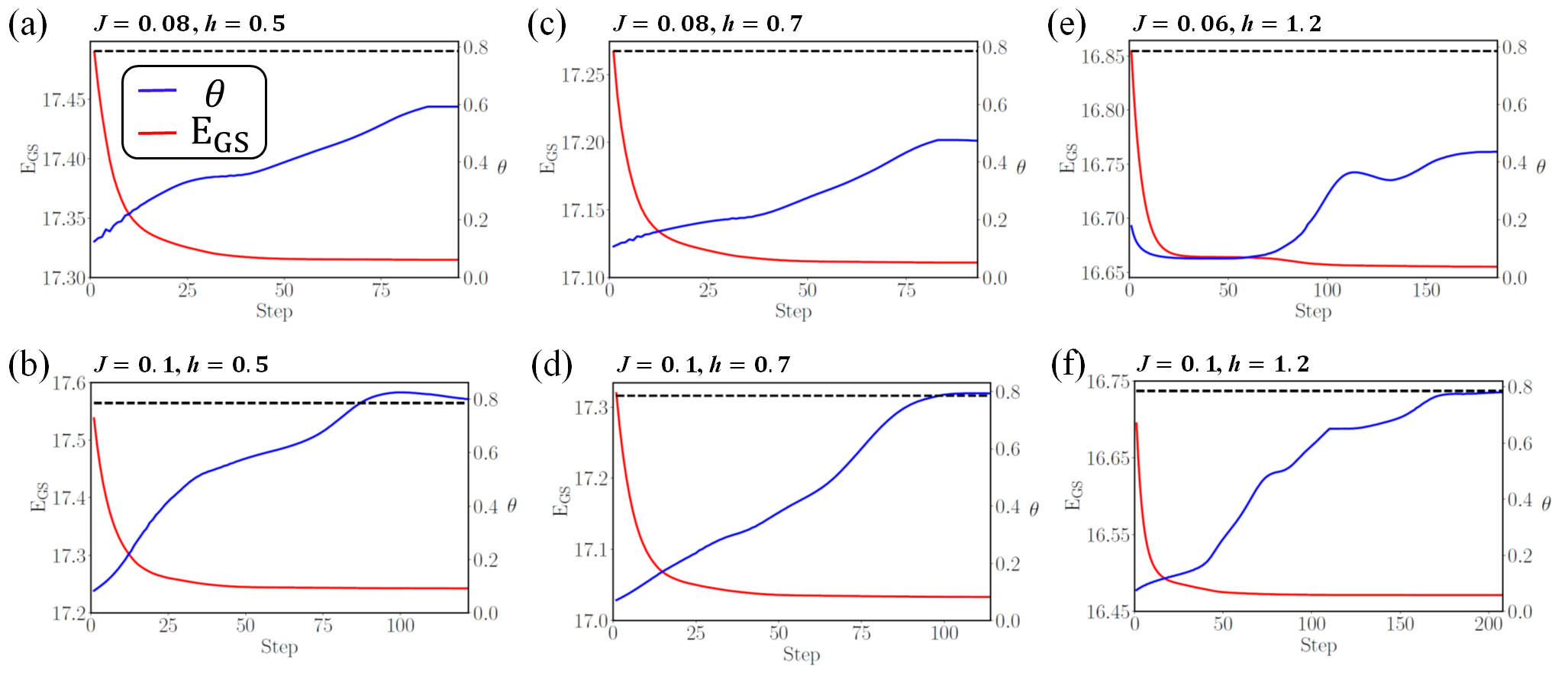}
\caption{Plots of ground state energy~($E_\text{GS}$) and azimuthal angle~($\theta$) defined in Fig.~\ref{fig1}(a) during the optimization process with $D=4$ for points near~((a), (c), and (e)) and away from~((b), (d), and (f)) the $P_2$ dome. The dashed lines indicate the $\theta=\pi/4$ orientation, where the magnetization is expected to align if not the easy-axis softening. The issue of local minimum is manifest in the upper row while not for the trials in the lower row.}
\label{fig5}
\end{figure*} 

\begin{figure}[pbt]
\centering
	\includegraphics[width=1.0\columnwidth]{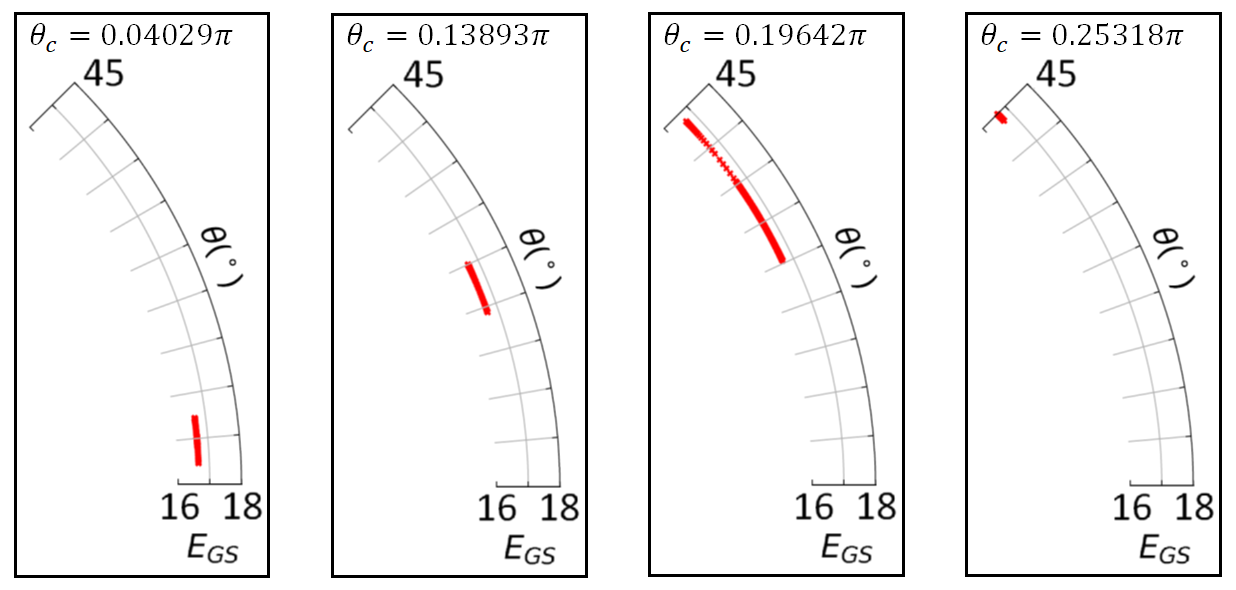}
\caption{$\theta$-$E_{\text{GS}}$ profiles for four distinct trials in Table~\ref{tab1} of $(J,h)=(0.06,1.2)$ are demonstrated. During the optimization process, as the intermediate $E_{\text{GS}}$ is smaller than the threshold value 16.656, the corresponding $\theta$ and energy are indicated with red dots in the one-eighth pie chart.}
\label{fig6}
\end{figure}

Moreover, a peculiar phenomenon happens as we approach the $P_2$ phase dome. 
We have discovered that the more we get close to its phase boundary, the more our calculation suffers from the local-minimum issue.
In Fig.~\ref{fig5} we have sampled several points near~((a), (c), and (e)) and away from~((b), (d), and (f)) the phase boundary and plotted the variation of GS energy~($E_\text{GS}$) and $\theta$ during the optimization process, all starting from the same random initial input.
As we can see, while at those points away from the phase boundary each trial converges to a canted ferromagnetic state with the converging angle $\theta_c\approx\pi/4$, close to the boundary the convergence flows to some local minima.
For a better comparison, in Table~\ref{tab1} we provide the detailed values of $E_\text{GS}$, $\theta_c$, and physical observables for the simulations of the points near the boundary.
For $h=0.5$ and $0.7$ we demonstrate two distinct results: values in the local and (nearly) global minima.
The globally minimal results are generated through the optimization starting from a converged ansatz nearby, which gives rise to a $\theta=\pi/4$ magnetic easy axis.
It is clear to see that despite some energy differences, $\Delta E$ is smaller than 0.005$\%$ of $E_\text{GS}$.
From the values of $M_z$ and $M_p$ we also realize that it is not simply the numerical artifacts when getting too close to the phase boundary, because $M_p$ is quite apparent and even larger than the 0.6 times of $M_z$.

For $h=1.2$, where the local-minimum issue is even more manifest, more data are shown with different directions of easy axis.
While the energy difference among them is even smaller than 0.0015$\%$ of $E_\text{GS}$, an ansatz giving rise to magnetic easy axis along the principal axis ($\theta_c/\pi=0.00018$) can also be detected.
For a better demonstration, in Fig.~\ref{fig6} we exhibit the $\theta$ of ansatz during the optimization process when the final convergence is well approached for several trials in Table~\ref{tab1}.
From the values of $E_{\text{GS}}$ in the table, converging ansatz all give energies smaller than 16.656.
As a result, we set this number as the threshold and plot the $\theta$ of ansatz when its energy already drops below 16.656 during the optimization process.
From Fig.~\ref{fig6} we can see that despite a larger or smaller fluctuation of $\theta$ near the final convergence, different trials flow to different $\theta_c$ which suggests the issue of emerging local minima.
Overall, we see that the whole $E_{\text{GS}}<16.656$ profile almost cover all the $\theta\in[0,\pi/4]$ region.
We interpret the above observation from our numerical results as the ``softening'' of easy axis, because as we get closer to the phase boundary among ansatz giving different $\theta_c$ they possess nearly indistinguishable energy difference.

In fact, the appearance of local minima is a common issue for iPEPS calculation, especially when the energy gradient is very small near the global minimum of Hilbert space manifold, or it is near the phase boundaries of first-order transition.
Since the latter scenario of a discontinuous transition does not apply here, it is reasonable to deduce that near the phase boundary the convergence toward the globally minimal point becomes more difficult with the gradient decent algorithm.
This implies an emergent phenomenon near the transition points, which corresponds to the insight of field theory and we will elucidate this point in the next section.

\section{\label{discussion}Interpreting the iPEPS Results}

We have demonstrated in the previous section that the easy axes tend to be smeared out near the critical points based on the iPEPS simulation.
This can be understood as the local-minimum issue in the manifold of GS energy.
Borrowing the knowledge from the well-established field theory, however, we realize that those local minima amount to an energy continuum due to the emergence of continuous symmetry on the critical points, justifying our iPEPS results.

\subsection{\label{HCB}Low-energy effective theory}

To unveil the cause of this local-minimum issue, let us first try to gain more understanding from the perspective of the lattice model by proposing an effective theory.
From Fig.~\ref{fig1}(c) we can clearly see that there is a large energy gap that separates $|\psi_1 \rangle$ and $|\psi_2\rangle$ from the rest of the states.
Therefore, when $J$ is enough to overcome the rather small energy difference between $E_{\psi_1}$ and $E_{\psi_2}$, a phase transition can be triggered.
In the vicinity of phase boundary~(or inside the $P_1$ and $P_2$ phases), we can project out the rest states other than $|\psi_1 \rangle$ and $|\psi_2\rangle$ on each site
\begin{equation}
\begin{aligned}
H^{\text{eff}}&=PHP\\
&=P(-J\sum_{\langle i,j \rangle}\vec{S}_i \cdot \vec{S}_j)P+\sum_{i,n}E_n|\psi_n^i\rangle\langle\psi_n^i|\\
&=H^{\text{eff}}_{J}+H^{\text{eff}}_{K-h},
\end{aligned}
\label{HamilFO}
\end{equation}
where $P=\Pi_i(1-\sum_{m=3}^5|\psi_m^i\rangle\langle\psi_m^i|)$ is the projection operator and $n=1,2$. Because we have the following relations
\begin{equation}
\begin{aligned}
&\langle \psi_1 |S^z|\psi_1 \rangle =2N^2(1-\alpha^2),\\
&\langle \psi_2 |S^z|\psi_2 \rangle =1,\\
&\langle \psi_1 |S^+|\psi_2 \rangle =\langle \psi_2 |S^-|\psi_1 \rangle =2N,
\end{aligned}
\label{algebra}
\end{equation}
while all the other terms are zero, the first term becomes
\begin{equation}
\begin{aligned}
H^{\text{eff}}_{J}=&-J\sum_{\langle i,j \rangle}\vec{\tau}_i \cdot \vec{\tau}_j,
\end{aligned}
\label{HamilFO1}
\end{equation}
with
\begin{gather}
 \tau^+=(\tau^-)^\dagger
 =
  \begin{bmatrix}
   0 &
   2N \\
   0 &
   0 
   \end{bmatrix},
 \tau^z
 =
  \begin{bmatrix}
   2N^2(1-\alpha^2) &
   0 \\
   0 &
   1 
   \end{bmatrix}.
\label{tau}
\end{gather}
With further elaboration, Eq.~(\ref{HamilFO1}) can be expressed as
\begin{equation}
\begin{aligned}
H^{\text{eff}}_{J}=&-t_J\sum_{\langle i,j \rangle}(\sigma^+_i \sigma^-_j + \sigma^-_i \sigma^+_j) +V_J\sum_{\langle i,j \rangle}\sigma^z_i \sigma^z_j -B_J\sum_i\sigma^z_i,
\end{aligned}
\label{HamilFO2}
\end{equation}
where $t_J=2JN^2$, $V_J=-J\gamma_1^2$, and $B_J=4J\gamma_1\gamma_2$ with $\gamma_1=N^2(1-\alpha^2)-\frac{1}{2}$ and $\gamma_2=N^2(1-\alpha^2)+\frac{1}{2}$; $\sigma$ stands for the Pauli matrices~(see Appendix~\ref{HCB2}). 
$H^{\text{eff}}_{J}$ represents a hard-core bosonic~(HCB) Hamiltonian with an attractive potential under a magnetic field.

Since effectively $H^{\text{eff}}_{K-h}$ is equivalent to an auxiliary ``field'' that discriminates $|\psi_1 \rangle$ and $|\psi_2\rangle$~(telling us where we are close to the first or second phase boundary), near the vicinity of phase boundary the properties of $C$ phase can be well described by $H^{\text{eff}}_{J}$.
Through the direct evaluation we learn that $|t|\gg|V|$ for the parameters of interest here; therefore, a superfluid condensation is favorable. 
More importantly, Eq.~(\ref{HamilFO2}) has a global $U(1)$ symmetry, meaning that after the spontaneous symmetry breaking its free energy does not change by an acquisition of an extra phase factor~($e^{i\phi}$) to the order parameter.
When the $U(1)$ symmetry is present, it forms a energy continuum where off-diagonal magnetization is free to align along any direction that spans the x-y plane.
Consequently, it leads to the freedom in choosing $\theta$ for the magnetic moment, and makes the energies all possess similar values~(degenerate right on the critical points).
This explains the reason why our simulation suffers from the local-minimum issue near the phase boundary during the numerical simulation.

The above-mentioned scenario becomes exact only when we can completely ignore the effect from $|\psi_3 \rangle$ to $|\psi_5\rangle$, suggesting that a sufficiently large energy gap is required and the tunneling~($J$) should be small~$(\lesssim|E_1-E_2|)$.
This is the reason why we still observe the $\theta=\pi/4$ canted state deep inside the $C$ phase, because there $J$ is much larger than $|E_1-E_2|$ and thus higher-energy states come into the play, fixing the magnetic easy axes along the tetragonal directions.
Moreover, as $h$ is small the energy gap also decreases~(Fig.~\ref{fig1}(c)); therefore the picture described by Eq.~(\ref{HamilFO2}) loses some credibility and the emergent phenomenon is less manifest, which also adheres to our observation to the numbers shown in Table~\ref{tab1}.

\subsection{The insight from field theory}

\subsubsection{\label{phi4theo}Classical field theory}

In the previous section we have unveiled the fact that the $U(1)$ symmetry which does not originate from our model gradually emerges approaching the phase boundary.
We then need to check whether it becomes exact right on the critical points.
This is feasible because there the correlation length diverges and thus those higher-energy terms causing the anisotropy, which we neglect in Eq.~(\ref{HamilFO}), become truly irrelevant~\cite{PhysRevB.16.1217, PhysRevB.29.5250, PhysRevB.61.3430}, while it should stay relevant in any distance away from the critical points~\cite{PhysRevB.11.478, PhysRevB.106.094424}.
To see whether this is true, we will extend the previous discussion by means of the field theory.
The critical behavior of $O(n)$ model with cubic anisotropy has firstly been extensively studied through the momentum-space RG for a LGW Hamiltonian composed of the $n$-vector model plus a diagonal quartic field term~\cite{PhysRevB.8.4270}.
Such LGW Hamiltonian possesses the following form
\begin{equation}
\begin{aligned}
{\cal H}=&\int d^dx \Big\{ \frac{1}{2}[(\nabla \phi)^2+r\phi^2]+u\phi^4+v\sum_{i=1}^{n}(\phi_i)^4 \Big\},
\end{aligned}
\label{phi4}
\end{equation}
where $(\nabla \phi)^2=\sum_{i=1}^{n}(\nabla \phi_i)^2$ and $\phi^2=\sum_{i=1}^{n}(\phi_i)^2$. $d$ stands for the real-space dimension and $n$ reflects the component of field $\phi$.
As the standard measure for a field Hamiltonian up to the quartic field terms, the physics as $d<4$ should be treated in a perturbative way with the $\epsilon=4-d$ expansion to pin down the fixed point for the critical behavior.
The RG flow diagrams have been first generated in Ref.~\cite{book.aharony} and then become a textbook material~\cite{book.chaikin, book.kardar, book.cardy}.
The latest flow diagrams can be seen in Ref.~\cite{PhysRevB.105.104101} with expansion up to the $6^{th}$ order in $\epsilon$.
An important feature of the RG flow is that there is a critical value for its component, $n_c$, above which the stable fixed point would change~\cite{PhysRevB.8.4270}.
When $n<n_c$ the critical point is dominant by the $O(n)$ universality class, while as $n>n_c$ the cubic fixed point becomes the stable one.

While the physical scenario does not change with higher-order $\epsilon$-expansion, the value of $n_c$ will be altered when a further calculation is conducted. 
To date, the most reliable results from $\epsilon$-expansion studies up to six-loop calculation~\cite{PhysRevB.61.15136, ADZHEMYAN2019332}, Monte Carlo~\cite{PhysRevB.84.125136}, and a very recent bootstrapping method~\cite{PhysRevD.104.105013} all indicate $2.85<n_c(d=3)<3$, suggesting that the Heisenberg cubic ferromagnet~($n=3$) in three dimensions should undergo a cubic phase transition when the temperature is dropping down to its Curie point.
An important message from the RG prediction also tells us that when $n<n_c$, the transition is governed by a continuous symmetry which is absent in the original Hamiltonian, meaning that the anisotropy term becomes irrelevant right at the transition point. 
This fact signifies an emergence of continuous symmetry on the phase boundary.
Such phenomenon has already been found in previous studies for the antiferromagnetic transverse-field frustrated Ising model~(TFFIM), whose component is equal to one, in (2+1) dimensions with LGW approach or Monte Carlo simulation~\cite{PhysRevB.63.224401,PhysRevB.68.104409,PhysRevB.96.115160}.

Since Eq.~(\ref{Hamil}) contains a Zeeman term, we need to re-consider its effective LGW Hamiltonian in the field theory.
Recall that in Section~\ref{MF} we have emphasized that for our system $M_z$ is irrelevant due to the non-zero longitudinal magnetic field.
As a result, for our phase transition of interest, $P \rightarrow C$, we only have two related components, and it leads to $n=2$ in the field theory interpretation.
The corresponding order parameter is $M_p=\sqrt{M_x^2+M_y^2}$, where $M_{\alpha}=\langle S^{\alpha}\rangle$ and $\alpha=x,y$.
Therefore we can write down its corresponding free energy up to the fourth power~\cite{DOMANSKI1988306}
\begin{equation}
\begin{aligned}
{\cal F}={\cal F}_0+\frac{r}{2}M_p^2+uM_p^4+v(M_x^4+M_y^4),
\end{aligned}
\label{FE}
\end{equation}
where ${\cal F}_0={\cal F}_0(M_z,h)$.
We ignore the higher-order terms in the free energy because they will only result in some quantitative changes~\cite{Phys.StatusSolidiB129}.
As Landau and Ginzburg have argued, near the continuous transition point Eq.~(\ref{FE}) can be re-expressed with coupled fields, leading to the free energy density
\begin{equation}
\begin{aligned}
f=f_0+\frac{1}{2}[(\nabla \phi)^2+r\phi^2]+u\phi^4+v\sum_{i=1}^{2}(\phi_i)^4,
\end{aligned}
\label{FE2}
\end{equation}
where $\phi^2=\sum_{i=1}^{2}(\phi_i)^2$ and $(\nabla \phi)^2=\sum_{i=1}^{2}(\nabla \phi_i)^2$, representing the kinetic term.
Because of ${\cal F}=-k_BT\text{log}{\cal Z}$, where
\begin{equation}
\begin{aligned}
{\cal Z}=\int {\cal D}\vec{\phi}\,e^{-\beta {\cal H}[\vec{\phi}]}
\end{aligned}
\label{PF}
\end{equation}
is the partition function, we can see that Eq.~(\ref{FE2}) turns out corresponding to the integrand in Eq.~(\ref{phi4}) with only two components instead of three.
A recent work by Venus utilizing RG upon 2D XY model with fourfold anisotropy reveals that the finite-size Kosterlitz-Thouless~(KT) transition takes place with weak anisotropy, while a crossover to the Ising criticality happens as the strength of anisotropy increases~\cite{PhysRevB.105.235440}.
The existence of KT transition also implies that there is a quasi-ordered phase between the high-temperature paramagnetic phase and the low-temperature ferromagnetic phase~\cite{PhysRevE.101.060105}.
As a result, starting from the $C$ phase near the phase boundary in zero temperature~(recall that the anisotropy needs to be weak), our 2D lattice model can feasibly host two KT transitions in finite temperature.
We will leave the further investigation for future studies.

\subsubsection{\label{QPT}Quantum phase transition}

In zero temperature our MFA and iPEPS calculations reveal two phase boundaries of quantum phase transition~(QPT).
In the general form of quantum field theory~(QFT), we write down the corresponding Lagrangian density
\begin{equation}
\begin{aligned}
{\cal Z}=&\int {\cal D}\Psi{\cal D}\Psi^*\,\text{exp} \Bigl( -\int d\tau d^dx {\cal L}\Bigl),\\
{\cal L}=&K_1\Psi^*\frac{\partial\Psi}{\partial\tau}+K_2|\frac{\partial\Psi}{\partial\tau}|^2-\frac{1}{2}[|\nabla \Psi|^2+r|\Psi|^2]\\
&-u|\Psi|^4-v\sum_{i=1}^{2}(\Phi_i)^4,
\end{aligned}
\label{lagran}
\end{equation}
where $\Psi=\Phi_1+i\Phi_2$ stands for the order parameter and $\tau$ is the imaginary time. 
Neglecting the anisotropy term~($v=0$), Eq.~(\ref{lagran}) represents the standard model for the dilute Bose gas~\cite{book.sachdev}.
Under such scenario, it is well-known that as long as $K_1\neq0$, $K_2$ becomes irrelevant after rescaling; therefore, we end up with a dynamical critical exponent $z=2$.
With $d=2$, the QFT lies right on the upper critical dimension for the quartic field term. 
Moreover, at the multicritical point where the particle-hole symmetry is present, $K_1=0$ and thus $z=1$~\cite{PhysRevB.40.546}.
Notably, as $z=1$ the QFT becomes Lorentz-invariant, meaning that the imaginary time can be treated as an independent extra dimension.

\begin{figure*}[pbt]
\centering
	\includegraphics[width=2.0\columnwidth]{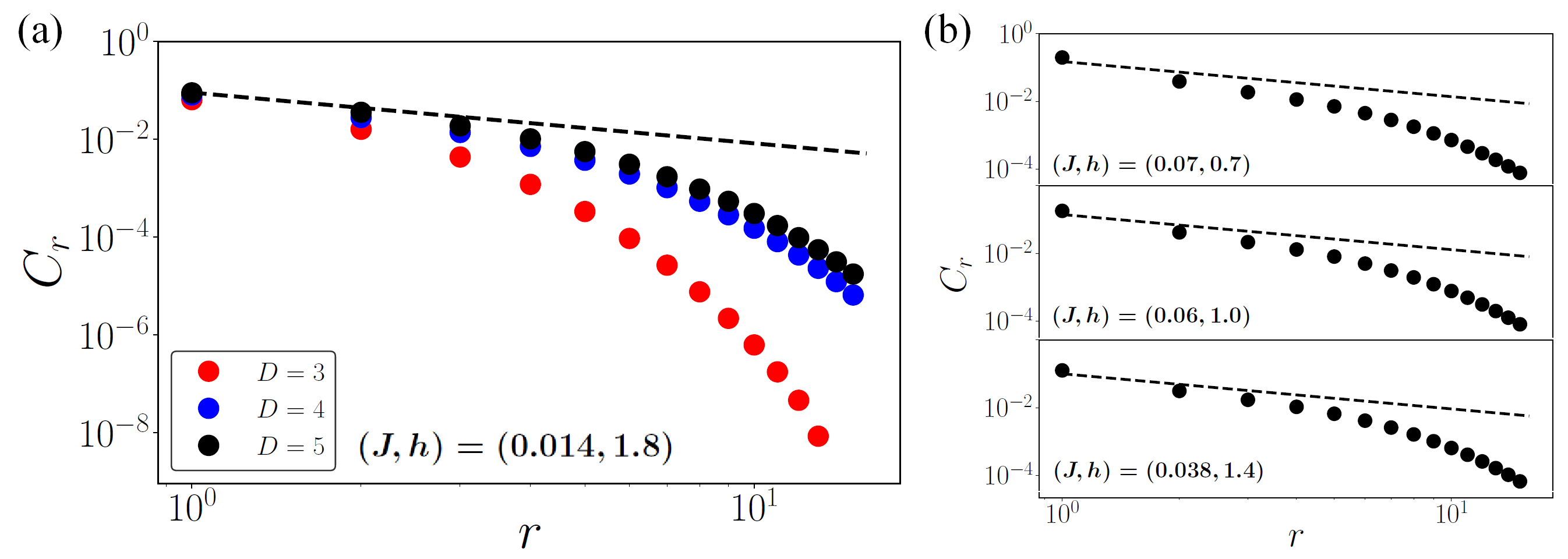}
\caption{Correlation functions in log-log scale at different points on the estimated phase boundary. Results of $(D,\chi)=(5,50)$, $(4,70)$, and $(3,100)$ are provided for $(J,h)=(0.014,1.8)$ in (a), while the samplings for other points with $(D,\chi)=(5,50)$ are also shown in (b). Dashed lines represent the power-law decay of correlation function in 3D XY criticality with $\eta=0.038$.}
\label{fig7}
\end{figure*} 

With $v\neq0$, it is difficult to deduce the property of the criticality on the current stage, although we believe that our scenario is akin to the diluted Bose gas quantum criticality with perturbation discussed in Ref.~\cite{PhysRevLett.76.4412}.
There, the authors demonstrated that the critical lines~(phase boundaries) are conformally invariant while their intersection~(multicritical point) is not~($(J,h)=(0,2)$ in Fig.~\ref{fig2}).
For our scenario, on the other hand, we could also rely on our numerical tool to provide us with further information.
In studying the quantum criticality, some important information can be extracted from the connected correlation function
\begin{equation}
\begin{aligned}
C_r=\sum_\alpha \langle S^\alpha_0S^\alpha_{r\hat{x}} \rangle -M_\alpha^2,
\end{aligned}
\label{correlationf}
\end{equation}
where $\alpha=x,y$. How to obtain the numerical correlation function is explained in Appendix~\ref{scaling}.
Right on the critical point the correlation length~($\xi$) diverges and $C_r$ scales as
\begin{equation}
\begin{aligned}
C_r\sim r^{-d+2-\eta},
\end{aligned}
\label{corrp}
\end{equation}
where $\eta$ is the critical exponent of correlation function.
Despite the fact that the finite-$D$ iPEPS is never able to capture the divergence of $\xi$, within the short range we can still approximate $\eta$ within a reasonable size of bond dimension~\cite{PhysRevB.97.174408, 10.21468/SciPostPhys.10.1.012}.

In Fig.~\ref{fig7}(a) we provide the log-log plots of $C_r$ for $(J,h)=(0.014,1.8)$ with three different $D$. Note that $(D,\chi)=(5,50)$ already reaches the maximal set-up within our machine's capacity.
Dashed lines represent the function of Eq.~(\ref{corrp}) with $\eta=0.038$, which is the critical exponent for the 3D XY universality class.
We can see that as $D$ increases the simulated curve of $C_r$ moves toward the dashed line and for $D=5$, a nice fitting along that line for small $r$ can be seen.
We have also sampled some other points in Fig.~\ref{fig7}(b) and all of the results seem to indicate the liability that the QPTs on distinct points belong to the same universality class.
Nevertheless, the numerical correlation functions exhibit an exponential decay in the longer range, and this is because of the construction of iPEPS that accords to the area law.
That is the reason why we demonstrate the trend for different $D$ in Fig.~\ref{fig7}(a), showing that by increasing the bond dimension the power-law decay can be better captured.
Similar analysis can also be seen in Refs.~\cite{10.21468/SciPostPhys.10.1.012, PhysRevB.97.174408}.

It is important to note that since our numerical correlation function demonstrates a nice fitting with the 3D XY universality class, which is a Lorentz-invariant criticality, we can be convinced that the $K_1$ term in Eq.~(\ref{lagran}) disappears.
As a result, our Lagrangian density is equivalent to the free energy density in Eq.~(\ref{FE2}), with an extra dimension for the imaginary time.
The XY universality class also roots for the picture of emergent $U(1)$ symmetry on the critical points.
In sum, our numerical simulation using iPEPS not only indicates the emergent phenomenon but also provides evidence for the property on the critical points, which is not easy to study with analytical approaches.
A further confirmation for the universality class is expected and left for future works.

\section{Conclusion}

In this work we study a lattice spin model composed of the ferromagnetic Heisenberg term and a cubic anisotropy under a nonzero magnetic field.
Utilizing the 2D iPEPS tensor network ansatz, we obtain a more accurate phase diagram than the one by MFA in the quantum regime.
Despite a four-fold degeneracy of magnetization in the canted phase resulting from the anisotropy, our discovery indicates an easy-axis softening near the critical phase boundary, signifying an emergent $U(1)$ symmetry.
Combining the perspective of field theory and iPEPS results, the $U(1)$ symmetry feasibly becomes exact right on the phase boundary and the criticality belongs to the 3D XY universality class.

Our study also indicates that for such spinful systems MFA can already capture the correct phases, suggesting that the phenomenological models built for real spinful semiconductors such as the Zener's model are reliable. 
This is not surprising because oftentimes $S=2$ is high enough and very close to the $S\to \infty$ classical limit.
Nevertheless, by exploiting the capability of advanced numerical ansatz, we can attain the consistency of our quantum many-body model with field-theory prediction for its critical behavior.
Since the zero-temperature quantum critical point is prone to induce a remnant effect in the finite temperature, our work also implies some interesting physics with thermal fluctuation near the QPT point.

Moreover, our study demonstrates the fact that the iPEPS tensor network ansatz is still useful for the gapless critical phases.
Unlike in one dimension where critical phases can be probed by the multiscale entanglement renormalization ansatz~(MERA)~\cite{PhysRevLett.99.220405}, it is not easy to extend similar construction in higher dimensions, making iPEPS almost the only choice in true 2D.
However, its construction also suggests that to probe the gapless phases one might need a very large bond dimension, whose computational cost easily goes beyond the capacity of the classical machine.
This fact makes the iPEPS ansatz with executable but small bond dimension~($D<10$) worrisome for the gapless phases.
In our study, on the other hand, we have shown that even with a small $D$ the critical behavior can already be well approximated.
Thanks to the unconventional lattice model we consider here whose criticality concerns an emergence of continuous symmetry from the insight of RG, the easy-axis softening reflects the emergent $U(1)$ symmetry and surprisingly, this phenomenon manifests with $D\leq 5$.
We believe that our work, joining in the group of previous works of importance studying the scaling at critical points~\cite{PhysRevX.8.031030, PhysRevX.8.031031, PhysRevX.8.041033, PhysRevLett.129.200601}, provides another solid example demonstrating the utility of iPEPS in a different but desirable way.
Since the emergence of continuous symmetry has been a widely seen phenomenon in numerous many-body systems, such as the well-known deconfined criticality~\cite{doi:10.1126/science.1091806, doi:10.1126/science.aad5007, PhysRevResearch.2.033459, PhysRevLett.125.257204, PhysRevLett.128.087201}, we believe that our results can bring attention to the usage of tensor network ansatz for its research, besides the QMC or analytical studies.
By enlarging the bond dimension, on the other hand, we also expect that the emergence of $U(1)$ symmetry can be better captured.
Therefore, how to apply iPEPS more efficiently for our model will be one of our future considerations.

We would like to point out two promising directions for future works.
First, a further confirmation of the critical nature is appreciable.
While the QMC calculation is hindered for this lattice model, a direct derivation of its QFT from the path integral representation or adopting the conformal field theory could be the liable candidate in validating our current scope.
Second, informed by the recent work~\cite{PhysRevB.105.235440} that a 2D XY model with anisotropy can host a finite-temperature KT transition, to study the finite-temperature behavior of our model is also favorable.
Since we have a lattice spin model, it is more suitable in describing the micro-mechanism of a 2D ferromagnetic thin film.
By adopting the thermal state purification through iPEPS and infinite projected entangled-pair operator~(iPEPO), its behavior in finite temperature can be probed~\cite{PhysRevB.99.035115, PhysRevLett.122.070502}, providing us with further information.

At last, we would like to briefly mention the connection of our study to the real-world materials, especially the layered diluted magnetic semiconductor~(DMS)~\cite{RevModPhys.86.187}.
It is believed that DMS plays an important role in designing new devices for future spintronics~\cite{nat.mat.9.952} and it is important to manipulate the magnetic easy axes for such purpose.
In the earlier studies, through applying an external electric field the magnetic easy axes can be shifted due to the altering of carrier concentration led by the field~\cite{nature.408.944, doi:10.1126/science.1086608, nature.455.515, nat.nano.10.209}.
Our study here shows that at low temperature~(where the quantum nature manifests) and near the critical points, easy axes disappear and the behavior is akin to a $U(1)$ magnet~(here we focus on the easy axes within the x-y plane).
Materials with cubic anisotropy such as (Ga,Mn)As~\cite{PhysRevB.97.184403}, (In,Mn)As~\cite{doi:10.1063/1.1885190}, or (Ga,Mn)P~\cite{PhysRevB.75.214419, PhysRevB.81.205210} then become ideal platforms in examining our proposal.
Although the Mn$^{2+}$ ions are often characterized in spinful $S=5/2$ state~\cite{RevModPhys.86.187}, it does not alter the field theory interpretation and thus our conclusion still holds.
Future studies for $S=5/2$ many-body Hamiltonian with cubic anisotropy will also be of interest.
In practice, a feasible way to investigate the easy-axis softening is to detect the angular dependence of ferromagnetic resonance field under a perpendicular external magnetic field~\cite{PhysRevB.75.214419}, which has, to the best of our knowledge, not been done by any group yet.
In sum, the easy-axis softening introduced here unveils a new possibility in manipulating the magnetism for spinful semiconductors.

In conclusion, although Eq.~(\ref{Hamil}) is just the simplest spin model of Heisenberg ferromagnet with cubic anisotropy, it already contains a huge potential which has not been fully exploited and thus merits future studies.
For example, in 2D monolayers~\cite{NatRevPhys.1.538, PhysRevB.97.184403} or quantum wells~\cite{PhysRevLett.79.511} the uniaxial anisotropy plays an important role and thus needs to be considered, too.
Frequently, in real-world materials the orbital angular momentum is not zero and thus the spin-orbital coupling, as well as the potential Jahn-Teller effect, could exist~\cite{refId0, EDATHUMKANDY2022169738}.
Therefore, to describe the true behavior of a practical system a more complex model needs to be built and considered. 
We hope that the results we presented here stimulate further research on the spin-anisotropy, which has been brought within reach by recent technical advancement.

\section{\label{Acknowledgement}Acknowledgement}
Authors appreciate decent comments and discussion from Eun-Gook Moon, Jun Takahashi, and Tsuyoshi Okubo. Part of the calculation was conducted in the Supercomputer Center of ISSP, the University of Tokyo.
W.-L.T. and H.-Y.L. are supported by National Research Foundation of Korea under the grant numbers NRF-2020R1I1A3074769.
%
%
H.-Y.L. is also supported by Basic Science Research Program through the National Research Foundation of Korea~(NRF) funded by the Ministry of Education~(2014R1A6A1030732).
X.L. is grateful to the support of the Global Science Graduate Course~(GSGC) program of the University of Tokyo.
S.R.G. and N.K. are supported by JSPS KAKENHI Grant No. JP19H01809. 
H.-K.W. is supported by JQI-NSF-PFC~(NSF Grant No. PHY-1607611).

\appendix

\section{\label{qmc}The difficulty in applying QMC}

As many of the physicists would have agreed, when handed with a many-body model on the lattice, QMC is one of the most reliable numerical tools that we can think of~\cite{book.kawashima, book.becca}. 
It has demonstrated its strength in many aspects~\cite{RevModPhys.73.33}, such as probing the various deconfined quantum criticality which is still an active research field~\cite{doi:10.1126/science.1091806, doi:10.1126/science.aad5007, PhysRevResearch.2.033459, PhysRevLett.125.257204, PhysRevLett.128.087201}.
However, due to the vicious sign problem~\cite{PhysRevB.41.9301} it hinders the utility of QMC upon many correlated systems of more interest.
The origin of sign problem comes from the negative weight of a configuration which could happen in a system with frustration or a fermionic system.
In general, if the negative sign of our Hamiltonian, $-H$, has negative off-diagonal elements, it is likely to cause the sign problem.

For our lattice model~(Eq.~(\ref{Hamil})), we have known that the Heisenberg term causes no sign problem in the square lattice, nor the Zeeman term. 
For the $K$ term, is can be expanded by the $S=2$ operators and the result is
\begin{equation}
\begin{aligned}
&-\sum_{\alpha}(S^\alpha)^4=
\begin{bmatrix}
-21 & & & & -3\\
 & -18 & & &\\
 & & -24 & &\\
 & & & -18 &\\
-3 & & & &-21\\
\end{bmatrix}.
\end{aligned}
\label{Kexp}
\end{equation}
One can see that we have negative off-diagonal terms and thus negative weights of some configuration might appear.
However, in the conventional world-line QMC, summing up all possible world-line configurations for the partition function, the periodic boundary condition along the direction of imaginary time forces the action of $K$ term being an even number.
As a result, we can dodge the bullet of sign problem.

Nevertheless, an obscure difficulty will happen for our model.
As mentioned above, in QMC one needs to determine the weight for each world-line configuration to construct the partition function, and the standard way is by putting the configuration through a Markov process, where the configuration goes through a series of local flips~\cite{doi:10.1143/JPSJ.73.1379}.
In our model, however, we have a magnetic anisotropic term~($K$) which could flip the $|2\rangle$ state into $|-2\rangle$ at each site.
Although the Heisenberg term~($J$) also flips the spin, it demands at least four times of acting $J$ term to connect $|2\rangle$ and $|-2\rangle$ states.
This would make the configurations hardly evolve because some processes, such as creating $|2\rangle$ at one site while annihilating it at another site, now become very complicated during the evolution and it results in a world-line configuration that is hard to equilibrate, while such processes are doomed to happen in the $C$ phase.
Such difficulty would lead to either the freezing of configuration or an extraordinarily long computational time.
Even considering the more efficient loop update or worm algorithm, this obstacle still remains and thus the improvement is limited.
So far, we are not aware of any QMC algorithm that could resolve this issue.

\section{\label{scaling}Transition points by iPEPS}

\begin{figure}[pbt]
\centering
	\includegraphics[width=1.0\columnwidth]{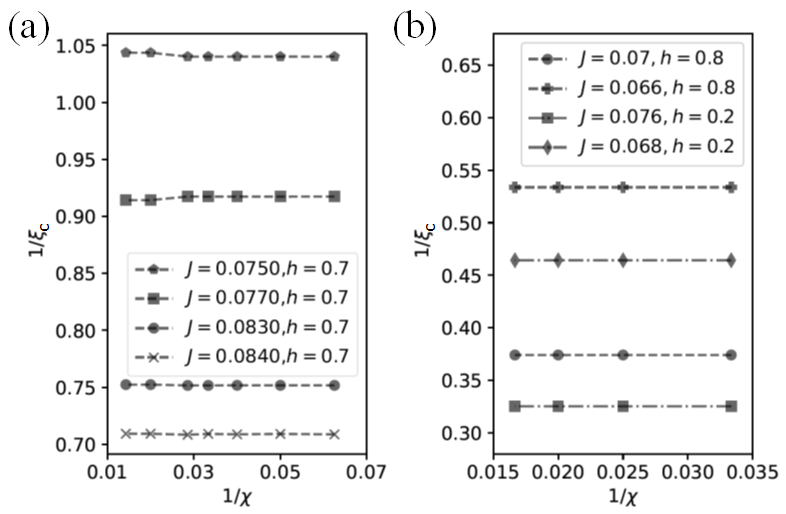}
\caption{We plot the characteristic $\xi$ obtained by Eq.~(\ref{trans_M}) for different $\chi$ in (a) $D=3$ and (b) $D=4$. From the data we can see that the variance of value in $\xi$ along $\chi$ is negligible.}
\label{fig8}
\end{figure} 

\begin{figure}[pbt]
\centering
	\includegraphics[width=1.0\columnwidth]{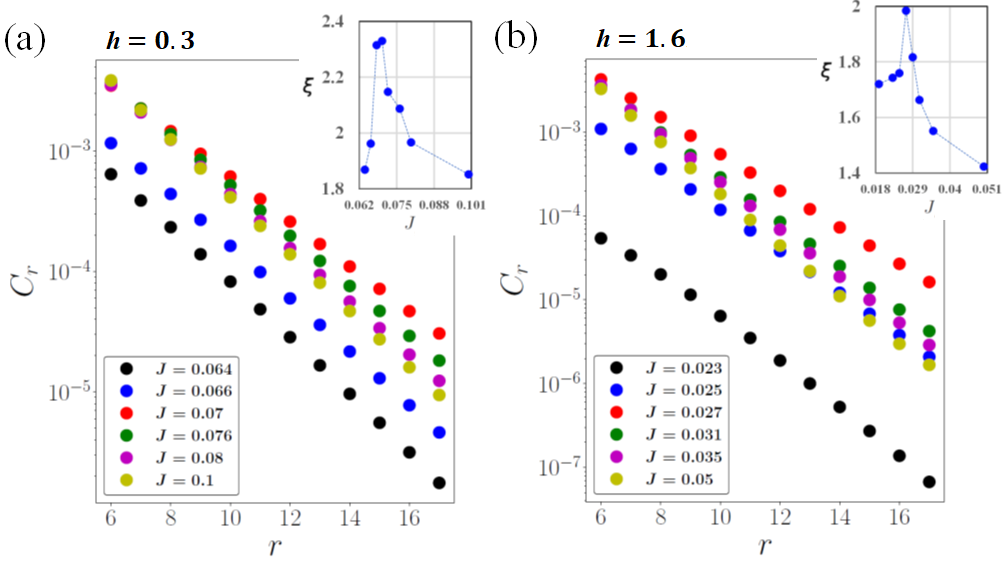}
\caption{The log-$r$ plots for $C_r$~($D=4$, $\chi=70$) along the vertical cuts at (a) $h=0.3$~($J_c\sim0.07$) and (b) $h=1.6$~($J_c\sim0.027$). Inset in each plot shows the estimated $\xi$ which is equal to the inverse minus slope of log($C_r$) in the long range for different $J$.}
\label{fig9}
\end{figure}

While iPEPS is an effective tensor network ansatz in two dimensions, its limitation lies on the available size of bond dimension.
Despite the fact that by utilizing the global symmetries of tensor network ansatz we are able to enlarge the bond dimension~\cite{PhysRevB.83.115125, PhysRevB.83.125106, PhysRevX.8.031031, PhysRevB.97.174408, 10.21468/SciPostPhys.10.1.012}, because of the lacking of desirable symmetry in our model it does not seem to be feasible and thus our calculation is constrained to smaller bond dimensions.
For our study using full tensors, the maximum setup which is affordable for our machine is $(D,\chi)\approx(5,50)$.

Because of the limitation, in this work we only provide the phase boundary estimated by finite $D$ without extrapolation to $D\to\infty$.
However, from Fig.~\ref{fig2} we also see that the estimated transition points barely change after $D=4$, fortifying the reliability of our phase diagram.
On the other hand, it is well-known that finite-$D$ iPEPS tends to over-emphasize the values of order parameter.
Without extrapolation a discontinuous behavior is likely captured by iPEPS calculation, blurring the accurate transition points.
Consequently, in this work we adopt the correlation length, $\xi$, as the indicator of continuous phase transition.

Since the finite-$D$ iPEPS can only capture the gapped phase, the estimated $\xi$ can never diverge even on the critical point.
However, near the continuous phase boundary we can still witness the ``rising up'' feature of $\xi$, despite the fact that it never goes to the infinity~\cite{PhysRevX.8.031030}.
To estimate $\xi$, one way we can follow is to construct the transfer matrix after the ansatz has converged and the ratio of its two leading eigenvalues represents the information of characteristic correlation length $\xi_c$~\cite{PhysRevB.97.174408, 10.21468/SciPostPhys.10.1.012}
\begin{equation}
\begin{aligned}
\xi_c=-\frac{1}{\text{log}|\lambda_1/\lambda_0|},
\end{aligned}
\label{trans_M}
\end{equation}
where the eigenvalues follow the descending order $|\lambda_0|>|\lambda_1|>|\lambda_2|\cdots$.
However, for the finite-$D$ iPEPS we cannot ignore the effect from the third~(or even higher order) leading eigenvalue, and thus an effective extrapolation formula has been proposed~\cite{PhysRevX.8.041033}.
Unfortunately, as shown in Fig.~\ref{fig8} without adopting a very large $\chi$, the estimated $\xi_c$ barely changes and thus we are not able to apply the extrapolation formula.
Fig.~\ref{fig8} also conveys the message that within the available values of $\chi$, the results of simulation are scarcely altered.

Instead of probing $\xi_c$ from the transfer matrix, in this article we estimate the values of correlation length from the correlation function, $C_r$, because the observables already converge well with smaller $\chi$~\cite{PhysRevX.8.031030, PhysRevX.8.031031}.
$C_r$ can be obtained by constructing the reduced density matrix as the following
\begin{equation}
\rho(r)=
\begin{diagram}
\draw[color=black!60, very thick](2, 0.5) rectangle (1, -0.5); \draw (1.5, 0) node {$C_1$};
\draw[color=black!30, line width=1mm] (2, 0) -- (2.5, 0); \draw[color=black!60, very thick](3.5, 0.7) rectangle (2.5, -0.5); \draw (3, 0) node {$T_1$};
\draw[color=black!30, line width=1mm] (1.5, -0.5) -- (1.5, -1); 
\draw[color=blue!60, line width=1mm] (3, -0.5) -- (3, -1); 
\draw[color=black!30, line width=1mm] (3.5, 0) -- (4, 0); 

\draw[color=black!60, very thick](2, -1) rectangle (0.8, -2); \draw (1.5, -1.5) node (X) {$T_4$};
\draw[color=blue!60, line width=1mm] (2, -1.5) -- (2.5, -1.5); \fill[color=red!60, very thick](3, -1.5) circle (0.5);
\draw[color=blue!60, line width=1mm] (3.5, -1.5) -- (4, -1.5);
\draw[color=black!30, line width=1mm] (1.5, -2) -- (1.5, -2.5); 
\draw[color=blue!60, line width=1mm] (3, -2) -- (3, -2.5); 
\draw[color=red!60] (3.4, -1.2) -- (3.9, -0.8);
\draw[color=red!60] (2.1, -2.2) -- (2.6, -1.8);

\draw[color=black!60, very thick](2, -2.5) rectangle (1, -3.5); \draw (1.5, -3) node {$C_4$};
\draw[color=black!30, line width=1mm] (2, -3) -- (2.5, -3); \draw[color=black!60, very thick](3.5, -2.5) rectangle (2.5, -3.7); \draw (3, -3) node {$T_3$};
\draw[color=black!30, line width=1mm] (4, -3) -- (3.5, -3);

\draw [black, thick] (9.5,1) to [square left brace ] (9.5,-4);
\draw [black, thick] (7.5,-4) to [square left brace ] (7.5,1);
\draw (11, 1) node {$(r-1)$};
\draw (5.5, -1.5) node {$\ldots$};
\draw (11.5, -1.5) node {$\ldots$};
\draw[color=black!30, line width=1mm] (15, 0) -- (14.5, 0); \draw[color=black!60, very thick](16, 0.5) rectangle (15, -0.5); \draw (15.5, 0) node {$C_2$};
\draw[color=black!30, line width=1mm] (15.5, -0.5) -- (15.5, -1); 
\draw[color=blue!60, line width=1mm] (15, -1.5) -- (14.5, -1.5); \draw[color=black!60, very thick](15, -1) rectangle (16.2, -2); \draw (15.5, -1.5) node (X) {$T_2$};
\draw[color=black!30, line width=1mm] (15.5, -2) -- (15.5, -2.5); 
\draw[color=black!30, line width=1mm] (15, -3) -- (14.5, -3); \draw[color=black!60, very thick](16, -2.5) rectangle (15, -3.5); \draw (15.5, -3) node {$C_3$};
\draw[color=black!30, line width=1mm] (13, 0) -- (13.5, 0); \draw[color=black!60, very thick](14.5, 0.7) rectangle (13.5, -0.5); \draw (14, 0) node {$T_1$};
\draw[color=blue!60, line width=1mm] (13, -1.5) -- (13.5, -1.5); \fill[color=red!60, very thick](14, -1.5) circle (0.5);
\draw[color=black!30, line width=1mm] (13, -3) -- (13.5, -3); \draw[color=black!60, very thick](13.5, -2.5) rectangle (14.5, -3.7); \draw (14, -3) node {$T_3$};
\draw[color=blue!60, line width=1mm] (14, -1) -- (14, -0.5); 
\draw[color=blue!60, line width=1mm] (14, -2) -- (14, -2.5); 
\draw[color=red!60] (14.4, -1.2) -- (14.9, -0.8);
\draw[color=red!60] (13.1, -2.2) -- (13.6, -1.8);

\draw[color=black!30, line width=1mm] (7.5, 0) -- (8, 0); \draw[color=black!60, very thick](9, 0.7) rectangle (8, -0.5); \draw (8.5, 0) node {$T_1$};
\draw[color=black!30, line width=1mm] (9.5, 0) -- (9, 0);
\draw[color=blue!60, line width=1mm] (7.5, -1.5) -- (8, -1.5); \fill[color=red!60, very thick](8.5, -1.5) circle (0.5);
\draw[color=blue!60, line width=1mm] (9.5, -1.5) -- (9, -1.5);
\draw[color=black!30, line width=1mm] (7.5, -3) -- (8, -3); \draw[color=black!60, very thick](9, -2.5) rectangle (8, -3.7); \draw (8.5, -3) node {$T_3$};
\draw[color=black!30, line width=1mm] (9.5, -3) -- (9, -3);
\draw[color=blue!60, line width=1mm] (8.5, -1) -- (8.5, -0.5); 
\draw[color=blue!60, line width=1mm] (8.5, -2) -- (8.5, -2.5); 
\end{diagram},
\label{eq:cr}
\end{equation}
where
\begin{equation}
\begin{diagram}
\draw[color=blue!60, line width=1mm] (1.5, -0.3) -- (1.5, -0.8);
\draw[color=blue!60, line width=1mm] (0.5, -1.3) -- (1, -1.3); \fill[color=red!60, very thick](1.5, -1.3) circle (0.5);
\draw[color=blue!60, line width=1mm] (2, -1.3) -- (2.5, -1.3);
\draw[color=blue!60, line width=1mm] (1.5, -1.8) -- (1.5, -2.3);
\end{diagram}
=
\begin{diagram}
\draw[line width=1mm] (1.1, 1.0) -- (1.5, 0.4);
\draw[line width=1mm] (0.5, 0) -- (1.3, 0); \draw[color=red!60, very thick](1.8, 0) circle (0.5); \draw (1.8, 0) node {$A$};
\draw[line width=1mm] (2.3, 0) -- (3.1, 0);
\draw[line width=1mm] (2.1, -0.4) -- (2.5, -1);
\draw[color=red!60] (1.8, -0.5) -- (1.8, -2);
\draw[line width=1mm] (1.1, -1.5) -- (1.5, -2.1);
\draw[line width=1mm] (0.5, -2.5) -- (1.3, -2.5); \draw[color=red!60, very thick](1.8, -2.5) circle (0.5); \draw (1.8, -2.5) node {$A^\dagger$};
\draw[line width=1mm] (2.3, -2.5) -- (3.1, -2.5);
\draw[line width=1mm] (2.1, -2.9) -- (2.5, -3.5);
\end{diagram}
\;\;\;\;\text{and}\;\;\;\;
\begin{diagram}
\draw[color=blue!60, line width=1mm] (1.5, -0.3) -- (1.5, -0.8);
\draw[color=blue!60, line width=1mm] (0.5, -1.3) -- (1, -1.3); \fill[color=red!60, very thick](1.5, -1.3) circle (0.5);
\draw[color=blue!60, line width=1mm] (2, -1.3) -- (2.5, -1.3);
\draw[color=blue!60, line width=1mm] (1.5, -1.8) -- (1.5, -2.3);
\draw[color=red!60] (1.9, -1.0) -- (2.4, -0.6);
\draw[color=red!60] (0.6, -2.0) -- (1.1, -1.6);
\end{diagram}
=
\begin{diagram}
\draw[line width=1mm] (1.1, 1.0) -- (1.5, 0.4);
\draw[line width=1mm] (0.5, 0) -- (1.3, 0); \draw[color=red!60, very thick](1.8, 0) circle (0.5); \draw (1.8, 0) node {$A$};
\draw[line width=1mm] (2.3, 0) -- (3.1, 0);
\draw[line width=1mm] (2.1, -0.4) -- (2.5, -1);
\draw[color=red!60] (1.8, -0.5) -- (1.8, -1);
\draw[color=red!60] (1.8, -1.5) -- (1.8, -2);
\draw[line width=1mm] (1.1, -1.5) -- (1.5, -2.1);
\draw[line width=1mm] (0.5, -2.5) -- (1.3, -2.5); \draw[color=red!60, very thick](1.8, -2.5) circle (0.5); \draw (1.8, -2.5) node {$A^\dagger$};
\draw[line width=1mm] (2.3, -2.5) -- (3.1, -2.5);
\draw[line width=1mm] (2.1, -2.9) -- (2.5, -3.5);
\end{diagram}.
\label{eq:phystrace}
\end{equation}
After contracting $\rho(r)$ with $S^\alpha$~($\alpha=x,y$), we obtain $\langle S^\alpha_0S^\alpha_{r\hat{x}} \rangle$. Then $C_r$ is evaluated by Eq.~(\ref{correlationf}).
In the longer range where the exponentially decay of $C_r$ takes place, it scales as
\begin{equation}
\begin{aligned}
C_r\sim e^{-r/\xi},
\end{aligned}
\label{corrp2}
\end{equation}
and thus, the inverse minus slope of log($C_r$)-$r$ plot for $r\gg1$ provides us with the estimated value of $\xi$.
According to the previous studies $\xi$ and $\xi_c$ should share similar values after a well estimation~\cite{PhysRevB.97.174408}.

In Fig.~\ref{fig9} we demonstrate the log($C_r$)-$r$ plots for the correlation function at different points along the vertical cut at $h=0.3$~($1.6$).
We can see that the red dots decrease the slowest and it represents $J=0.07$~($J=0.027$).
In the insets we show the estimated $\xi$ and indeed there is a summit around the two points; consequently the critical points can be approximated.
By repeating the same procedure for different vertical cuts, we end up producing the phase diagram in Fig.~\ref{fig2}.

\section{\label{HCB2}Effective hard-core bosonic model}

In Section~\ref{HCB} we present an effective Hamiltonian of HCB from the low-energy $H^{\text{eff}}_J$ and we will show the derivation here. First, re-write Eq.~(\ref{HamilFO1}) into
\begin{equation}
\begin{aligned}
H^{\text{eff}}_J=&-2JN^2\sum_{\langle i,j \rangle}\sigma^+_i \sigma^-_j+\sigma^+_j \sigma^-_i\\
&-J\sum_{\langle i,j \rangle}(\gamma_1\sigma^z_i+\gamma_2\mathds{1}_i)(\gamma_1\sigma^z_j+\gamma_2\mathds{1}_j),
\end{aligned}
\label{HamilFO3}
\end{equation}
with $\gamma_1=N^2(1-\alpha^2)-\frac{1}{2}$, $\gamma_2=N^2(1-\alpha^2)+\frac{1}{2}$, and \(\mathds{1}\) stands for the identity matrix. After expanding Eq.~(\ref{HamilFO3}), we obtain 
\begin{equation}
\begin{aligned}
H^{\text{eff}}_J=&-2JN^2\sum_{\langle i,j \rangle}\sigma^+_i \sigma^-_j+\sigma^+_j \sigma^-_i-J\gamma_1^2\sum_{\langle i,j \rangle}\sigma^z_i\sigma^z_j\\
&-4J\gamma_1\gamma_2\sum_i\sigma^z_i,
\end{aligned}
\label{HamilFO4}
\end{equation}
where the $\mathds{1}_i\mathds{1}_j$ term has been ignored since it only contribute a constant energy. By assigning $t_J=2JN^2$, $V_J=-J\gamma_1^2$, and $B_J=4J\gamma_1\gamma_2$, we obtain the Hamiltonian in Eq.~(\ref{HamilFO2}).

\bibliography{draft}

\end{document}


\title{Studies for quantum ferromagnet with cubic anisotropy with and beyond the mean-field theory}

\author{Wei-Lin Tu}
\affiliation{Division of Display and Semiconductor Physics, Korea University, Sejong 30019, Korea}

\author{S. R. Ghazanfari}
\affiliation{Institute for Solid State Physics, University of Tokyo, Kashiwa, Chiba 277-8581, Japan}

\author{Huan-Kuang Wu}
\affiliation{Department of Physics, Condensed Matter Theory Center and Joint Quantum Institute, University of Maryland, College Park, MD 20742, USA}

\author{Hyun-Yong Lee}
\email{hyunyong@korea.ac.kr}
\affiliation{Division of Display and Semiconductor Physics, Korea University, Sejong 30019, Korea}
\affiliation{Department of Applied Physics, Graduate School, Korea University, Sejong 30019, Korea}
\affiliation{Interdisciplinary Program in E$\cdot$ICT-Culture-Sports Convergence, Korea University, Sejong 30019, Korea}

\author{Naoki Kawashima}
\email{kawashima@issp.u-tokyo.ac.jp}
\affiliation{Institute for Solid State Physics, University of Tokyo, Kashiwa, Chiba 277-8581, Japan}

\date{\today}

\maketitle

\section{\label{section1}Section I}

\begin{figure}[!hbt]
\centering
	\includegraphics[width=1.0\columnwidth]{FigA1.png}
\caption{}
\label{figA1}
\end{figure}

The iPEPS tensor network ansatz is an effective numerical method for dealing with various quantum problems in two dimension \cite{PhysRevLett.101.250602, PhysRevB.84.041108, PhysRevLett.113.046402, NatRevPhys.1.538}. It has many merits that some other numerical techniques do not have, such as its attainability to the thermodynamic limit and freedom from the vicious sign problem in quantum Monte Carlo. This ansatz contains two parts, the bulk tensors and environmental tensors for achieving the infinite size. Technically, the number of bulk tensors can be freely chosen, but a better choice will allow it to be able to reflect the real space modulation for the ground state (GS). Thus, as shown in Fig. \ref{figA1}(a), in this work we apply a $2\times 2$ unit cell for the bulk, meaning that there are four different rank-5 bulk tensors from $a_1$ to $a_4$. For each tensor it has four virtual bonds (bonds that connect nearby tensors), capturing the entanglement between site to site, with bond dimension $D$. The remaining one leg of each tensor, connecting $a_n$ and $a_n^\dagger$, is the physical bond with dimension $d$, equal to the dimension of local Hilbert space.

\begin{table}
\centering
\caption{}
\begin{tabular}{cccccc}\hline\hline
            & \begin{tabular}[c]{@{}c@{}}~~2$\times$2~~\\\end{tabular} & ~~3$\times$2~~ & ~~4$\times$2~~ & ~~3$\times$3~~  \\ 
\hline
$B=0.4$($z$-AFM) & -0.837(6)                                               & -0.79(1)  & -0.837(6)  & -0.82(0)     \\ 
\hline
$B=0.6$(SS)       & -0.841(1)                                               &  -0.80(2)  & -0.841(1)  & -0.79(2)    \\
\hline
$B=1.5$(Mott)       & -1.288(3)                                               &  -1.18(2)  & -1.288(2)  & -1.17(6)    \\
\hline
$B=2.5$(SSF)       & -1.863(8)                                               &  -1.858(0)  & -1.863(6)  & -1.843(1)    \\
\hline\hline
\end{tabular}
\label{tabA1}
\end{table}
%



\bibliography{draft}